\documentclass[journal]{IEEEtran}
\ifCLASSINFOpdf
\else
\fi

\usepackage{amsthm}

\usepackage{moreverb}
\usepackage{multirow}
\usepackage{algorithmic}

\usepackage{textcomp}
\usepackage{mathptmx} 
\usepackage{times} 
\usepackage{enumerate} 
\usepackage{verbatim} 

\usepackage{graphics} 
\usepackage{epsfig} 
\usepackage{amsmath} 
\usepackage{amssymb}  
\usepackage{amsfonts}
\usepackage{epstopdf}
\usepackage{graphicx}
\usepackage{subfigure}
\usepackage{tabularx}
\usepackage{array}
\usepackage{booktabs}
\usepackage{cite}
\usepackage{float}
\usepackage{dblfloatfix}
\usepackage{lipsum}
\RequirePackage{fixltx2e}
\usepackage{silence} 
\newtheorem{Lemma}{Lemma}
\newtheorem{Theorem}{Theorem}

\newtheorem{Assumption}{Assumption}
\newtheorem{Remark}{Remark}

\newtheorem*{Proof}{Proof}
\DeclareMathAlphabet{\mathcal}{OMS}{cmsy}{m}{n}
\DeclareSymbolFont{largesymbols}{OMX}{cmex}{m}{n}

\RequirePackage[normalem]{ulem} 
\RequirePackage{color} 

\def\BibTeX{{\rm B\kern-.05em{\sc i\kern-.025em b}\kern-.08em
		T\kern-.1667em\lower.7ex\hbox{E}\kern-.125emX}}

\begin{document}
	
\title{Segmented Model-Based Hydrogen Delivery Control for PEM Fuel Cells: a Port-Hamiltonian Approach}
	
\author{Lalitesh Kumar, Jian Chen$^*$, Chengshuai Wu, Yuzhu Chen, and Arjan van der Schaft
\thanks{Jian Chen and Yuzhu Chen are with the State Key Laboratory of Fluid Power and Mechatronic Systems, School of Mechanical Engineering, Zhejiang University, Hangzhou 310027,China. }
\thanks{Lalitesh Kumar is with the State Key Laboratory of Fluid Power and Mechatronic Systems, College of Control Science and Engineering, Zhejiang University, Hangzhou 310027, China.} 
\thanks{Chengshuai Wu is with the School of Automation Science and Engineering, Xi'an Jiaotong University, Xian 710049, China.}
\thanks{Arjan van der Schaft is with the Bernoulli Institute for Mathematics, Computer Science and Artificial Intelligence, University of Groningen, Groningen 9747 AG, the Netherlands.}
\thanks{This work is supported by the Key Research and Development Program of Zhejiang Province under Grant 2021C01098.}
\thanks{* Corresponding author.    Email: jchen@zju.edu.cn.}
	}


\maketitle

\begin{abstract}
This paper proposes an extended interconnection and damping assignment passivity-based control technique (IDA-PBC) to control the pressure dynamics in the fuel delivery subsystem (FDS) of proton exchange membrane fuel cells. The fuel cell stack is a distributed parameter model which can be modeled by partial differential equations (PDEs). In this paper, the segmentation concept is used to approximate the PDEs model by ordinary differential equations (ODEs) model. Therefore, each segments are having multiple ODEs to obtain the lump-sum model of the segments. Subsequently, a generalized multi-input multi-output lumped parameters model is developed in port-Hamiltonian framework based on mass balance to minimize the modeling error. The modeling errors arises due to the difference between spatially distributed pressures in FDS segments, and also due to the difference between the actual stack pressure and the measured output pressure of the anode. The segments interconnection feasibilities are ensured by maintaining passivity of each segment. With consideration of re-circulation and bleeding of the anode in the modeling, an extended energy-shaping and output tracking IDA-PBC based state-feedback controller is proposed  to control the spatially distributed pressure dynamics in the anode. Furthermore, a sliding mode observer of high order is designed to estimate the unmeasurable pressures in FDS with known disturbances. Performance recovery of output feedback control is accomplished with explicit stability analysis. The effectiveness of the proposed IDA-PBC approach is validated by the simulation results.
 \end{abstract}

\markboth{\textit{}}%
{}

\section{Introduction}

\IEEEPARstart{F}{uel} Cells have gained attentions during the last two decades due to its certain incomparable advantages such as exorbitant energy density as compared to the available traditional fuels, no contamination to the environment due to only water dumping at the outlet, low temperature operation, and higher efficiency \cite{Sharma1,Ogungbemi2}. Despite advantages mentioned above, the proton exchange membrane fuel cells (PEM-FCs) have certain pitfalls that impede the large-scale commercialization of fuel cells \cite{Dubau3}. Thus, there is a need to design the efficient control systems to tune the operating parameters in optimal way. In view of this, the detailed investigations have been done on parametric changes to discover the best operating parameters \cite{9146674}. In addition to tune the optimal parameters, the consumption of hydrogen has to be maximized in the anode flow channel. Keeping this in mind, the authors in \cite{LIU2020115110} have designed a nitrogen observer for anode flow channel to deal with varying loads of the PEM-FCs, and to maximize the hydrogen consumption at an excellent rate of 99 \%. The majority of research in the literature has concentrated on lumped parameter modeling of fuel cells, with little attention given to modeling with segmentation concept. The segmentation technique were used to address the problems of water management in \cite{Esmaili2020ModelBW}, the power management in \cite{4939358} and the fault tolerance in \cite{6043104}, the life cycle in \cite{WENG20103664}, and the thermal management in \cite{TOLJ201113105} for the PEM fuel cells. Furthermore, the authors in \cite{https://doi.org/10.1002/asjc.1827} addressed the control of required flow and concentration of hydrogen based on the segmentation of anode flow channel. They have designed a sliding mode controller with high gain observer, but the design process is very cumbersome and it costs complexity and computational burden. Besides this, the authors in \cite{CHEN20081179} and \cite{CHEN20111992} proposed a segmented, lumped, and calibrated model based on patterns in different flow fields. The model's partial differential equations are approximated in each segment based on the patterns developed in the anode and cathode flow channels. Later on, a nonlinear state-feedback controller was propounded to control the hydrogen delivery in the PEM-FCs \cite{HONG20171565}. In the following paragraph, we will discuss the port-Hamiltonian control framework and how it is incorporated to model and control the various aspects of PEM fuel cells.

The port-Hamiltonian (pH) based control design technique is developed in recent decade which in particular can be used as a geometrical framework to model the physical systems as a network model through the Dirac structure. It provides the unified mathematical framework to analyze, model, and control the systems of different physical domains for both distributed and lumped parameters \cite{10.5555/3285631}. The pH concept for the controller design has gained impetus in the last few years, in particular, interconnection and damping assignment passivity-based control (IDA-PBC) being the most popular approach to deal with the control design \cite{ORTEGA2002585}. However, in recent years, the shifted passivity control design has also been developed in pH framework \cite{WU2019595, 9127091}, which outperforms the IDA-PBC in few applications. The stabilization theory of IDA-PBC based pH systems have been dealt in \cite{CASTANOS20091611,WU2020109087, PANTELEY1998131}. Furthermore, in \cite{MONSHIZADEH2019108527}, the control actions and disturbances are applied to balance the power not to the flow variables and for preserving the pH structure, an observer is designed by solving the matching equations in \cite{8386663}. Besides, the stabilization of under-actuated mechanical systems are proposed in\cite{1024334} and an extension of IDA-PBC is proposed in \cite{DONAIRE2016118}. Because IDA-PBC is based on the energy-shaping concept, but there are limits on the conversion of energy and its transfer from one port to another \cite{9416793}. However, due to its energy-shaping attitude, the IDA-PBC-based control approach has significant advantages in energy conversion devices such as in speed regulation of the PM synchronous motors \cite{960344,832798}. In fuel cells systems, the smart energy management \cite{BENMOUNA201922467}, mitigation of fuel starvation \cite{HILAIRET20131097}, stabilization issues of DC micro-grid with a multi-phase fuel cell \cite{9032209}, and load maintenance for converters at different voltage levels in \cite{8818313} have been investigated by incorporating the IDA-PBC approach.

The thoroughgoing literature study reveals that on segmentation of fuel cells, very few works are available and mostly on the material research. Also, the segmented model for fuel delivery control in the pH framework has not received adequate attention from researchers yet, this motivated us to shorten such gaps. The research on energy shaping pH control has pragmatical significance for the complex nonlinear energy conversion systems, and it is worth mentioning here.  We have proposed an extended IDA-PBC-based energy shaping pH controller aiming to control the spatially distributed pressures in fuel delivery subsystem (FDS) of PEM fuel cell systems. Although many research works are available in the literature of fuel cells that involves pH control framework, but most of them are dealing with the power electronics aspect of fuel cells without considering the spatially distributed properties. To the best of authors' knowledge and belief, non of the work has been done on the investigation of chemical aspects of the PEM fuel cells using passivity based pH approach. The main contributions of this paper are encapsulated as follows: 
\begin{itemize}
	\item \emph{\textbf{}}The present paper proposes a generalised multi-input multi-output (MIMO) model of FDS in pH framework based on the segmentation concept to reduce the modelling error caused by the difference between spatially distributed pressures in the segments, as well as the modelling error caused by actual stack pressure and measured stack pressure. The passivity of each segment is maintained, as is the pH structure of each segment, ensuring the interconnection feasibility of the segments. \item \emph{\textbf{}}Based on IDA-PBC approach, an extended energy-shaping and output tracking state-feedback controller is proposed to control the spatially distributed pressure dynamics in the segmented FDS model of PEM fuel cells. \item \emph{\textbf{}}The unmeasurable pressures are estimated by a sliding mode observer of high order. Based on estimated states by the observer, an output feedback controller in pH framework is designed with performance recovery. Moreover, an apodictic analysis of stability of the closed-loop system is shown in this paper. 
\end{itemize}
The remaining part of the paper is organized as follows: Section \uppercase\expandafter{\romannumeral2} describes the mathematical model development. Section \uppercase\expandafter{\romannumeral3} reports the control design, segments interconnection feasibility and stability analysis. A high order sliding mode observer and performance recovery of output feedback is presented in section \uppercase\expandafter{\romannumeral4}. Finally, Section \uppercase\expandafter{\romannumeral5} contains results and their discussions and the work is concluded in  Section \uppercase\expandafter{\romannumeral6}.

\textbf{Nomenclature}: The following nomenclatures are considered in this paper: $A$ is area ($m^2$), $F$ is Faraday constant (c/mol), $I$ is fuel current (A), $I_n$ is the $n^{th}$ segment anode current, $\dot \eta$ is Molar flow rate (mol/s), $N_{fc}$ is Fuel cell number, $M$ is molar mass (g/mol), $\mathcal P$ is the pressure (Pa), $\mathcal P$ is atmospheric pressure (Pa), $R$ is ideal gas constant ($J/(mol.K)$), $T$ is temperature ($K$), and $V$ is the volume ($m^3$).

\textbf{Subscripts}: The following subscripts are used in the model: $a$ is for anode, $c$ is for cathode, $ani$ or $aqi$ indicates anode inlet orifices, $an$ or $aq$ is used for $n^{th}$ or $q^{th}$ anode segment, $\dot \eta_{ani}$ or $\dot \eta_{aqi}$, $bd$ indicate bleed orifice, $bl$ is for blower, $crn$ or $crq$ indicates $k^{th}$ or $q^{th}$ segments crossover process, $ht$ is for hydrogen tank, $q$ or $n$ indicates $q^{th}$ or $n^{th}$ segment, $s$ is for species, $v$ is for vapor, and $sm$ indicate the supply manifold.  

\section{Development of the Mathematical Models}
This section describes the dynamical nonlinear MIMO state-space mathematical model of the distributed parameters FDS. The model for re-circulation of the hydrogen is developed by considering the exhausted bleeding to ensure the flow of hydrogen to the anode channel with required rate and concentration. 
The anode channel of PEM fuel cell is partitioned into $n$ segments, where the first segment will receive input hydrogen flow, and the required output will come out through the $n^{th}$ segment, therefore, there will be $n-2$ segments between the input and the output segments. The schematic design of the segmented anode channel FDS is shown in Fig. 1. The following assumptions are taken for the development of model:
\begin{itemize}
	\item \emph{\textbf{}}The properties of parameters in each segment are uniform, not to vary with the segments \cite{CHEN202048}. \item \emph{\textbf{}}The relative humidity is maintained at more than 95\% in the anode flow channel \cite{HONG20171565}. 
	\item \emph{\textbf{}}The currents derivative with respect to time are provided with a bound with known finite system constants \cite{6919261}. 	\item \emph{\textbf{}}The temperature of anode channel is perpetuated to an acceptable limit by cooling mechanism in the loop \cite{HONG20171565}.	\item \emph{\textbf{}}The volume of each segments are same and the operation of cathode is normal. 
\end{itemize} 	 

\subsection{Modeling of Segmented Anode}
The characteristics of parameters in each segment are assumed to be uniform, according to the assumption. This results in mass conservation at each node (interconnection point of the two segments) of the segmented model, as well as reduced spatial variation of states in the FDS. 
According to the ideal gas law, the gases follow the relations $PV=\eta RT$, with all the variables having their usual physical meaning. Consequently, the required pressure control dynamics can be modeled as $\frac{dP}{dt} = \frac{RT}{V} \left[\frac{d\eta}{dt}\right]$. Taking conservation of mass into account with the laws of ideal gases, the pressure dynamics of nitrogen and hydrogen gases in FDS can be mathematically represented as follows \cite{HONG20171565}:
\begin{itemize}
	\item \emph{\textbf{The first segment:}}
	\begin{equation}
		\begin{split}
			\frac{d\mathcal{P}_{a1,H_2}}{dt}&=\frac{RT_a}{V_{a1}}\left[\dot {\eta}_{a1i,H_2}-\dot {\eta}_{a2i,H_2}-\dot {\eta}_{r1,H_2}-\dot {\eta}_{cr1,H_2}\right]\\ 
			\frac{d\mathcal{P}_{a1,N_2}}{dt}&=\frac{RT_a}{V_{a1}}\left[\dot {\eta}_{a1i,N_2}+\dot {\eta}_{cr1,N_2}-\dot {\eta}_{a2i,N_2}\right].\label{eq:1}
		\end{split} 
	\end{equation}
	\item \emph{\textbf{The $q^{th}$ segment:}}
	\begin{equation*} 
		\resizebox{0.925\hsize}{!}{$
			\frac{d\mathcal{P}_{aq,H_2}}{dt}=\frac{RT_a}{V_{aq}}\left[\dot {\eta}_{aqi,H_2}-\dot {\eta}_{a(q+1)i,H_2}-\dot {\eta}_{rq,H_2}-\dot {\eta}_{crq,H_2}\right]\\
			$}
	\end{equation*}
	\begin{equation}
		\frac{d\mathcal{P}_{aq,N_2}}{dt}=\frac{RT_a}{V_{aq}}\left[\dot {\eta}_{aqi,N_2}+\dot {\eta}_{crq,N_2}-\dot {\eta}_{a(q+1)i,N_2}\right].\  \label{eq:2}                     		
	\end{equation}
	\item \emph{\textbf{The $n^{th}$ segment:}}
	\begin{equation*} 
		\resizebox{0.9\hsize}{!}{$
			\frac{d\mathcal{P}_{an,H_2}}{dt} \!=\! \frac{RT_a}{V_{an}}[\dot{\eta}_{ani,H_2} \!-\! \dot{\eta}_{bd,H_2}  \!-\! \dot{\eta}_{rn,H_2} \!-\! \dot{\eta}_{crn,H_2} \!-\! \dot{\eta}_{bl,H_2}] \\
			$}
	\end{equation*}
	\begin{equation}
		\frac{d\mathcal{P}_{an,N_2}}{dt}=\frac{RT_a}{V_{an}}\left[\dot {\eta}_{ani,N_2}+\dot {\eta}_{crn,N_2}-\dot{\eta}_{bd,N_2}-\dot{\eta}_{bl,N_2}\right] \label{eq:3} \                      
	\end{equation}
\end{itemize}
with $q= 2,3,...,n-1$ and the molar flow rates of the gases 
$\dot {\eta}_{a1i,H_2}$, $\dot {\eta}_{aqi,N_2}$, $\dot {\eta}_{a(q+1)i,N_2}$, and $\dot {\eta}_{ani,N_2}$ with  $\dot {\eta}_{bd,H_2}$, $\dot {\eta}_{bd,N_2}$ can be obtained using the following orifice equations:
\begin{figure}[!t]\centering
	\includegraphics[width=8cm, height=5cm]{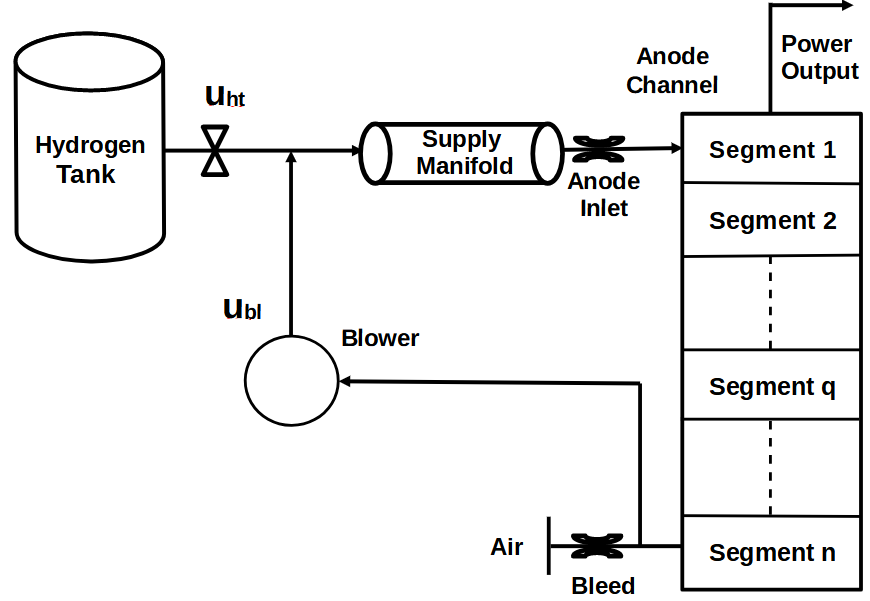}
	\caption{PEM Fuel Cell Segmented Model.}
	\label{PEM Fuel Schematic}
\end{figure}
\begin{itemize}
	\item \emph{\textbf{The first segment:}}
	\begin{equation}
		\begin{split}
			\dot{\eta}_{a1i,s} &= \frac{\alpha A_{ai}}{M_{ej}}(\mathcal{P}_{sm} - \mathcal{P}_{a1})\frac{\mathcal{P}_{sm,s}}{\mathcal{P}_{sm}}\\
			\dot{\eta}_{a2i,s} &= \frac{\alpha A_{a1}}{M_{a1}} (\mathcal{P}_{a1} - \mathcal{P}_{a2}) \frac{\mathcal{P}_{a1,s}}{\mathcal{P}_{a1}}.  \label{eq:4}
		\end{split}
	\end{equation}
	\item \emph{\textbf{The $q^{th}$ segment:}}
	\begin{equation}
		\begin{split}
			\dot{\eta}_{aqi,s} &= \frac{\alpha A_{a(q-1)}}{M_{a(q-1)}} (\mathcal{P}_{a(q-1)} - \mathcal{P}_{aq}) \frac{\mathcal{P}_{a(q-1),s}}{\mathcal{P}_{a(q-1)}}\\
			\dot{\eta}_{a(q+1)i,s} &= \frac{\alpha A_{aq}}{M_{aq}} (\mathcal{P}_{aq} - \mathcal{P}_{a(q+1)}) \frac{\mathcal{P}_{aq,s}}{\mathcal{P}_{aq}}. \label{eq:5}
		\end{split}
	\end{equation}
	\item \emph{\textbf{The $n^{th}$ segment:}}
	\begin{equation}
		\begin{split}
			\dot{\eta}_{ani,s} &= \frac{\alpha A_{a(n-1)}}{M_{a(n-1)}} (\mathcal{P}_{a(n-1)} - \mathcal{P}_{an}) \frac{\mathcal{P}_{a(n-1),s}}{\mathcal{P}_{a(n-1)}}  \\
			\dot{\eta}_{bd,s} &= \frac{\alpha A_{bd}}{M_{an}} (\mathcal{P}_{an} - \mathcal{P}_{0}) \frac{\mathcal{P}_{an,s}}{\mathcal{P}_{an}} \label{eq:6} 
		\end{split}
	\end{equation}
\end{itemize}
where $s = \{H_2, N_2\}$ and $M_{ej}=M_{sm}$. In \eqref{eq:1} - \eqref{eq:3}, $ \dot{\eta}_{r1,H_2}, \dot{\eta}_{rq,H_2}$, and $\dot{\eta}_{rn,H_2} $ indicates the rate of reaction of hydrogen in the first, $q^{th}$, and $n^{th}$ segments of the anode channel, $ \dot{\eta}_{cr1,H_2}, \dot{\eta}_{crq,H_2}$, and $\dot{\eta}_{crn,H_2} $ defines the rate of permeation of the hydrogen in $1^{st}$, $q^{th}$, and $n^{th}$ segments of the anode. Additionally, $ \dot{\eta}_{cr1,N_2}, \dot{\eta}_{crq,N_2}$, and $\dot{\eta}_{crn,N_2}$ symbolizes the rate of permeation of nitrogen in $1^{st}$, $q^{th}$, and $n^{th}$ segments of the anode channel  with $q = 2,3,...,n-1 $. 

\subsection{Membrane Gas Consumption and Crossover}
The consumed hydrogen in segments $1^{st}$, $q^{th}$, and $n^{th}$  has obtained by the law of conservation of proton, and can be represented mathematically as follows:
\begin{equation}
	\begin{aligned}
		&\dot {\eta}_{r1,H_2}=\frac{N_{fc} I_{1}}{2F},..., \dot {\eta}_{rq,H_2}=\frac{N_{fc} I_{q}}{2F},..., \dot {\eta}_{rn,H_2}=\frac{N_{fc} I_{n}}{2F}.\label{eq:7}\\
	\end{aligned}
\end{equation}
The currents in the different segments will be $I_1$, $I_q$, and $I_n$. The summation $I_1+... + I_q+... + I_n$ = $I$ represents the total current flowing into the anode channel for $q=2,3,...,n-1$.
By neglecting the $ H_2 $ and $ N_2 $ permeation of the membrane and crossover of the oxygen, the molar rates of the crossover for the first, the $q^{th}$, and the $n^{th}$ segments can be represented as
\begin{equation}
	\begin{aligned}
		&\dot {\eta}_{cr1}={t_1N_{fc}k_{cr}(\mathcal{P}_{s,h}-\mathcal{P}_{s,l})}\\ 
		&\dot {\eta}_{crq}={t_qN_{fc}k_{cr}(\mathcal{P}_{s,h}-\mathcal{P}_{s,l})},q=2,3,...,n-1 \\  
		&\dot {\eta}_{crn}={t_nN_{fc}k_{cr}(\mathcal{P}_{s,h}-\mathcal{P}_{s,l})} \label{eq:8}
	\end{aligned}
\end{equation}
where $k_{cr}$ indicates the rate of permeation provided by the manufacturer. Furthermore, $ t_{k}, k=1,...,q...,n $ denotes the volume proportion of the segments. However, $\mathcal{P}_{s,h}$ and $\mathcal{P}_{s,l}$ are the higher and lower pressure sides of the $ H_2 $ and $ N_2 $ gases, respectively.
\subsection{Supply Manifold Model}
The supply manifold consists of three components, namely the hydrogen storage tank, the anode flow channel, and the blower model. The gases that flow through the supply channels of the fuel cell usually follow the ideal gas law. Referring to the assumptions made in Section II, the relative humidity has to be maintained almost constant at a level of greater than 95\%, and for the constant presupposition of vapor pressure, the dynamics for $ H_2 $ and $ N_2 $ can be derived as following:
\begin{align}
	\frac{d\mathcal{P}_{sm,s}}{dt}&=\frac{RT_m}{V_{sm}}\left[\dot {\eta}_{ht,s}+\dot {\eta}_{bl,s}-\dot {\eta}_{a1i,s}\right] \label{eq:9}
\end{align}
where $s=[{H_{2},N_{2}}]$. The control of $\dot {\eta}_{ht,H_2}$ is done in a proportional concept by providing a bound to the flow rate as follows:
$\dot {\eta}_{ht,H_2}={u_{hl}\dot {\eta}_{ht,m}}, \dot {\eta}_{ht,N_2}=0$.
\subsection{Cathode and Blower Models}
According to the assumptions made in Section II,  the cathode is operating in its normal state without any complexity. The representation of the cathode model is similar to that proposed in \cite{HONG20171565}. The fraction of nitrogen concentration is as shown below:
\begin{equation}
	\begin{split}
		x_{c,N_2} = & \frac{3.16F\alpha A_{or}(\mathcal{P}_c-\mathcal{P}_0)(1-x_{c,H_2O})}{4F\alpha A_{or}(\mathcal{P}_c-\mathcal{P}_0)-0.79N_{fc}I(M_{N_2}-M_{O_2})} \\& + \frac{0.79FN_{fc}I(M_{O_2} + (M_{H_2O} - M_{O_2})x_{c,H_2O})}{4F\alpha A_{or}(\mathcal{P}_c-\mathcal{P}_0)-0.79N_{fc}I(M_{N_2}-M_{O_2})}
	\end{split}
	\label{eq:10}
\end{equation}
with $\alpha$ and $A_{or}$ are the coefficient and area of the orifices, $M_{O_2}$, $M_{H_{2}O}$, and $M_{N_2}$ are the molar mass of gases $O_2$, ${H_{2}O}$, and ${N_2}$. The variable $\mathcal{P}_c$ is defined as the cathode nominal pressure and it is provided by the manufacturer.
The model of the blower can be written as following \cite{HONG20171565}:
\begin{equation}
	\left\{
	\begin{aligned}
		& \dot{\eta}_{bl} = u_{bl}W_{bl,m}/M_{bl}, \Phi = \Phi_m(1- e^{\beta (\frac{\Psi}{\Psi_m} - 1)}), \\ 
		& W_{bl,m} = \Phi \rho_{bl} \frac{\pi}{4} d^2_{bl} U_{bl}, \Psi = C_p T_{bl,in}(\frac{\mathcal{P}_{bl,out}}{\mathcal{P}_{bl,in}})^{\frac{\gamma-1}{\gamma}}/(\frac{1}{2}U^2_{bl}).	
	\end{aligned}
	\right.
	\label{eq:11}
\end{equation}
In \eqref{eq:11}, $\rho_{bl}$, $d_{bl}$, and $U_{bl}$ indicates the density, diameter, and the speed of blade tip, respectively. The variables $M_{bl}$, $W_{bl,m}$, $\phi$, and ${\Psi}$ represent the molar mass, flow rate of the blower mass at maximum rotational speed, normalized flow rate of the blower, and a dimensionless quantity. The constants $\gamma$ and $(\frac{C_P}{C_V})$ are the specific heat and molar heat capacity ratios. Besides, $T_{bl}$ and $u_{bl}$ are the temperature and control signal to re-circulation loop, and $\Phi_m$, $\beta$, and $\Psi_m$ can be represented by the following polynomial equations:
\[
\Phi_m = \Sigma^4_{i=0} a_i M^i,\ \ \beta = \Sigma^2_{i=0} b_i M^i, \ \ \Psi_m =  \Sigma^5_{i=0} c_i M^i
\]
where $ M = U_{bl}/ \sqrt{\gamma R_a T_{bl,in}} $. 

\subsection{The Integrated State-Space Model Development}
The state variables are defined as $x_{1,H_2}\triangleq\mathcal{P}_{a1,H_2}$, $x_1 \triangleq \mathcal{P}_{a1}$, $x_{q,H_2}\triangleq\mathcal{P}_{aq,H_2}$, $x_q\triangleq\mathcal{P}_{aq}$, $x_{n,H_2}\triangleq\mathcal{P}_{an,H_2}$, $x_n\triangleq\mathcal{P}_{an}$, $x_{sm,H_2}\triangleq\mathcal{P}_{sm,H_2}$, $x_{sm}\triangleq\mathcal{P}_{sm}$. The pressures $\mathcal{P}_{a1}\triangleq\mathcal{P}_{a1,H_2}+\mathcal{P}_{a1,N_2}$, $\mathcal{P}_{aq}\triangleq\mathcal{P}_{aq,H_2}+\mathcal{P}_{aq,N_2}$,\ $\mathcal{P}_{an}\triangleq\mathcal{P}_{an,H_2}+\mathcal{P}_{an,N_2}$,\ and\ $\mathcal{P}_{sm}\triangleq\mathcal{P}_{sm,H_2}+\mathcal{P}_{sm,N_2}$,\ respectively. The nonlinear MIMO state-space model of the FDS is developed with the help of dynamical models derived in previous sebsection as follows: 
\begin{itemize}
	\item \emph{\textbf{The first segment:}}
	\begin{equation}
		\begin{split}
			&\dot x_{1,H_2}=a_{11}x_{1,H_2}+a_{17}x_{sm,H_2}+\zeta_{1,H_2}\\
			&\dot x_1=a_{21}x_{1,H_2}+a_{22}x_1+a_{24}x_2+a_{28} x_{sm}+\zeta_{1}.
		\end{split}\label{first segment:24}
	\end{equation}
	\item \emph{\textbf{The $q^{th}$ segment:}}
	\begin{equation}
		\begin{split}		
			&\dot x_{q,H_2}=a_{31}x_{{(q-1)},H_2}+a_{33}x_{q,H_2}+\zeta_{q,H_2}\\
			&\dot x_q=a_{42}x_{(q-1)}+a_{43}x_{q,H_2}+a_{44}x_q+a_{46}x_{(q+1)}+\zeta_{q}.
		\end{split}\label{qth segment:25}
	\end{equation}
	\item \emph{\textbf{The $n^{th}$ segment:}}
	\begin{equation}
		\begin{split}
			&\dot x_{n,H_2}=a_{53}x_{{(n-1)},H_2}+a_{55}x_{n,H_2}+G_{n,H_2}u_1+\zeta_{n,H_2}\\
			&\dot x_n=a_{64}x_{(n-1)}+a_{65}x_{n,H_2}+a_{66}x_n+G_{n}u_1+\zeta_{n}.
		\end{split}\label{nth segment:26}
	\end{equation}
	\item \emph{\textbf{The supply manifold:}}
	\begin{equation}
		\begin{split}
			\dot x_{sm,H_2}&=a_{77}x_{sm,H_2}+G_{sm,H_2}u_1+G_{sm,ht} u_2\\
			\dot x_{sm}&=a_{82}x_1+a_{88}x_{sm}+G_{sm}u_1+G_{sm,ht} u_2\\
		\end{split}\label{supply_manifold:27}
	\end{equation}
\end{itemize} 
with auxiliary variables $a_{ij}, i,j=1, 2,..., 8$ are introduced to reduce the mathematical complexity, in Appendix A.

\section{State-Feedback PH Control Design}
\subsection{PH Control Design Preambles}
To facilitate the control design, the compact state-space model in pH framework of the system \eqref{first segment:24} - \eqref{supply_manifold:27} is developed  as below \cite{10.5555/3285631, 960344}:
\begin{equation}
	\begin{split}
		&\dot x=\left[J(x)-R(x)\right]\frac{\partial H}{\partial x}(x)+G(x_{n,H_2}, y)u+\zeta\\
		&y=C\frac{\partial H}{\partial x}(x)
	\end{split}\label{segmented_model_pH_framework}
\end{equation}
where $	C=
{\begin{bmatrix}
		0 & 0 & 0 & 0 & 0 & 1 & 0 & 0\\
		0 & 0 & 0 & 0 & 0 & 0 & 0 & 1\\
\end{bmatrix} }$.
The state and input vector are defined as follows:
\begin{equation*} 
	\begin{split}
		x(t)=&\left[x_{1,H_2}, x_1,... x_{q,H_2}, x_q,..., x_{n,H_2}, x_n, x_{sm,H_2}, x_{sm}\right]^T\\
		u(t)=&\left[u_1, u_2\right]^T=\left[u_{bl}, u_{ht}\right]^T.
	\end{split}
\end{equation*}
The output variables are considered as the pressures of supply manifold and $n^{th}$ segment of anode outlet, $\mathcal{P}_{sm}$ and $\mathcal{P}_{an}$. Therefore, the output $y$ is defined as $
y(t)=\left[\mathcal{P}_{an},\ \mathcal{P}_{sm}\right]^T=\left[x_{n},\ x_{sm}\right]^T.
$
In \eqref{segmented_model_pH_framework}, the functions $G(x)$ and $\zeta$ are defined as given below:
\begin{equation*}
	\begin{split}
		&G(x)=
		{\begin{bmatrix}
				0 & 0 & 0 & 0 & G_{n,H_2} & G_{n} & G_{sm,H_2} & G_{sm}\\
				0 & 0 & 0 & 0 & 0 & 0 & G_{sm,ht} & G_{sm,ht}\\
		\end{bmatrix} }^T\\&
		\zeta=\left[\zeta_{1,H_2}, \zeta_{1},...,\zeta_{q,H_2}, \zeta_{q},...,\zeta_{n,H_2}, \zeta_{n}, 0, 0 \right]^T 
	\end{split}
\end{equation*}
where $G_{sm,ht} \triangleq \mu_{sm} \dot\eta_{ht,m}$, $G_{n,H_2} \triangleq \mu_n\left(\frac{W_{bl,m}}{M_{an}}\right)\left(\frac{x_{n,H_2}}{x_n}\right)$, $G_{n} \triangleq \mu_n\left(\frac{W_{bl,m}}{M_{an}}\right)$, $G_{sm,H_2} \triangleq \mu_{sm}\left(\frac{W_{bl,m}}{M_{an}}\right)\left(\frac{x_{n,H_2}}{x_n}\right)$, and $G_{sm} \triangleq \mu_{sm}\left(\frac{W_{bl,m}}{M_{an}}\right)$. The disturbances $\zeta$ is defined in Appendix A. 
The interconnection matrix $J(x)$  with skew-symmetric properties and dissipation matrix $R(x)$  with symmetric and positive semi-definite properties are calculated for the segmented PEM fuel cell model, as shown in the Appendix B.  The control input $u$ is defined in $\mathbb{R} ^m$, and $\zeta$ indicates the exogenous disturbance signals as a function of currents in different segments, sensor noise, and unmodeled structures. The Hamiltonian $H(x):\mathbb{R} ^n \rightarrow \mathbb{R}_+$ for the fuel cell system is defined as follows:
\begin{equation}
	\begin{split}
		&H=\frac{1}{2}k_{1,H_2} x_{1,H_2}^2+\frac{1}{2}k_1 x_1^2+...+\frac{1}{2}k_{q,H_2} x_{q,H_2}^2+\frac{1}{2}k_q x_q^2+\\&
		...	+\frac{1}{2}k_{n,H_2} x_{n,H_2}^2+\frac{1}{2}k_n x_n^2+\frac{1}{2}k_{sm,H_2} x_{sm,H_2}^2+\frac{1}{2}k_{sm} x_{sm}^2.
	\end{split}
	\label{energy_open-loop}
\end{equation}
The constants $k_{j, H_2}$ and $k_j$ are defined as following:
$k_{j,H_2} = k_j = \frac{I_j RT_a}{nF}$, with $j = 1, \dots, q, \dots, n$. For reducing the complexity the constants are taken as  $k_{j,H_2} = k_j =k_{sm,H_2}= k_{sm} = 1$.\\ 
In IDA-PBC approach, the main objective is to design a suitable state-feedback control in such a way that the dynamics of closed-loop system is of the port controlled Hamiltonian (PCH) form. The objective is accomplished by shaping the desired Hamiltonian $H_d$ via assigning the desired interconnection matrix $J_d$ and the damping matrix $R_d$. The PCH form of the closed-loop system  can be represented as \cite{ORTEGA2002585}: 
\begin{equation}
	\dot x=\left[J_d(x)-R_d(x)\right]\frac{\partial H_d}{\partial x}(x).\\
	\label{state-feedback_closed-loop}
\end{equation}
In \eqref{state-feedback_closed-loop}, the matrices $J_d$ and $R_d$ are skew-symmetric and symmetric positive semi-definite and are assigned in appendix B.
For finding the desired stable equilibrium point, the appropriate control action has to be designed such that $J_d(x)=J(x)+J_a(x)$, and $R_d(x)=R(x)+R_a(x)$.
If $\sigma(x): \mathbb{R}^n\rightarrow\mathbb{R}^m$ is the control function, $\Omega(x): \mathbb{R}^n\rightarrow\mathbb{R}^m$ is a vector, and $J_a(x)$, $R_a(x)$ are square matrices such that \cite{ORTEGA2002585}
\begin{equation}
	\resizebox{1\hsize}{!}{$
		\left[J_d(x)-R_d(x)\right]\Omega(x)+\left[J_a(x)-R_a(x)\right]\frac{\partial H}{\partial x}=G(x_{n,H_2}, y)\sigma(x)+\zeta\\
		$}\label{eq:16}
\end{equation}
at the desired equilibrium point $x=x_d$. The following conditions are to be followed from \cite{ORTEGA2002585}:
\begin{enumerate}
	\item Integrability Condition
		\begin{equation}
				\left(\frac{\partial\Omega}{\partial x}(x)\right)=\left[\frac{\partial\Omega}{\partial x}(x)\right]^T. \label{Integrability_Condition}
			\end{equation}
	\item Equilibrium assignment condition
	\begin{equation}
		\Omega(x_d)= -\nabla H(x_d).\label{Equilibrium_Assignment_Condition}
	\end{equation}
	\item Lyapunov stability condition
	\begin{equation}
		\nabla \Omega(x_d)>-\nabla^2 H(x_d).\label{Lyapunov_Stability_Condition}
	\end{equation}
\end{enumerate}
The point $x_d=arg$ $min$ $H_d(x)$ is a stable equilibrium point (locally). For achieving the asymptotic stability the largest invariant set of the closed-loop system  is included in $x \in \mathbb{R}^n \mid \left\{\nabla^T H_d(x) R_d(x)\nabla^T H_d(x)=0\right\}$ equals $\left\{x_d\right\}$. 
Under the described conditions above, the closed-loop PEM fuel cell will be a pH controlled system with the desired energy function $H_d(x)=H(x)+H_a(x)$ has a minimum locally at $x_d$ and $H_a(x)$ satisfies the relation $  \Omega(x)=\nabla H_a (x)$.
\begin{Remark}
	The system dynamics \eqref{state-feedback_closed-loop} achieves the desired trajectory $x=x_d$, which
	is represented as
	\begin{equation}
		\dot x_d=\left[J_d(x_d)-R_d(x_d)\right]\frac{\partial H_d}{\partial x}(x_d) = F_d(x_d)\nabla H_d(x_d)
		\label{state-feedback_closed-loop_at desired trajectory}
	\end{equation}
	at $x=x_d$ and $F_d(x)\triangleq J_d(x)-R_d(x)$. Also, the function $F_d(x) +F^T_d(x) \leq 0$ as $F_d(x) +F^T_d(x)=-2R_d(x)$, and the desired damping matrix $R_d(x) \geq 0$. 
\end{Remark}
\subsection{IDA-PBC Based Controller Design}
The output tracking error is defined as $e \triangleq y - y_d $ with $e=\left[e_{n}, e_{sm}\right]^T$ and $y_d=\left[x_{nd}, x_{smd}\right]^T$, where $x_{nd}$ and $x_{smd}$ are the desired anode output and supply manifold pressures. As a consequence of the controller design precept given in \eqref{eq:16}, the PDE's obtained are utilized to shape the Hamiltonian function as follows:
	\begin{equation*} \resizebox{1\hsize}{!}{$
			H_a= \ln\left(b\right)
			\left(\frac{-b\zeta_{1,H_2}}{\mu_1 m_{ej}\left(1-\frac{x_1}{x_{sm}}\right)\left\{\mu_1 \rho_1+\mu_1 m_{1}-\mu_1 m_1\left(\frac{x_2}{x_1}\right)\right\}}\right)-	F_1\left(\bar b\right) +	
			$}
	\end{equation*}
	\begin{equation}\resizebox{1\hsize}{!}{$
			\ln\left(c\right)	\left(\frac{-c\zeta_{q,H_2}}{\mu_q m_{(q-1)}\left(1-\frac{x_q}{x_{(q-1)}}\right)\left\{\mu_q\rho_q+\mu_q m_q\left(1-\frac{x_{(q+1)}}{x_q}\right)\right\}}\right)-	F_2\left(\bar c\right) + \varphi(x_{n,H_2}) 	\label{eq:36}
			$}
	\end{equation}
	with 
	$b=a_{17}x_{1,H_2} + a_{11}x_{sm,H_2}$, $\bar b=a_{17}x_{1,H_2} - a_{11}x_{sm,H_2}$, $c=a_{31}x_{q,H_2} + a_{33}x_{{(q-1)},H_2}$, and $\bar c=a_{31}x_{q,H_2} - a_{33}x_{{(q-1)},H_2}$, respectively.
	Furthermore, $\varphi(x_{n,H_2})$ is the function of $x_{n,H_2}$ which is absent in the PDE's solution. Since the derivative $\frac{\partial H_d}{\partial x_{n,H_2}}$ depends on only $x_{n,H_2}$, the polynomial function $\varphi\left({x_{n,H_2}}\right)$ is propounded as
	\begin{equation}
		\varphi\left({x_{n,H_2}}\right)=\frac{\beta_1}{2}x^2_{n,H_2}-\beta_2x_{nd,H_2}\left(x_{n,H_2}-x_{nd,H_2}\right) \label{eq:42}
	\end{equation}
	where $\beta_1$ and $\beta_2$ are positive constants. 
	\begin{Remark} The functions $F_1$, $F_2$, and $\varphi\left({x_{n,H_2}}\right)$ are differentiable and selected so as to follow the conditions \eqref{Equilibrium_Assignment_Condition} and \eqref{Lyapunov_Stability_Condition}. Also, the functions $\dot F_1(\bar b)$ and $\dot F_2(\bar c)$ are upper bounded.
	\end{Remark}
	\begin{Remark}
		The hydrogen pressure in $n^{th}$ segment follow the relation $x_{n,H_2} > x_{nd,H_2}$ and hence, $\varphi\left({x_{n,H_2}}\right)$ is bounded in a small domain of interest to make the convergence of it at the desired equilibrium point $x_{nd,H_2}$ as
		\begin{equation*}
			\dot\varphi\left(x_{n,H_2}\right) = \left(\beta_1 x_{n,H_2} - \beta_2 x_{nd,H_2}\right)\dot x_{n,H_2},
		\end{equation*}
		and with some algebraic deductions, it can be proved that
		\begin{equation*}\resizebox{1\hsize}{!}{$
				\dot \varphi\left({x_{n,H_2}}\right) \leq  \lvert \beta_ 1 x_{n,H_2} - \beta_2 x_{nd,H_2}\rvert\  \left(\Gamma_1 -\Gamma_2\right) \leq \triangle_{\varphi}\|\tilde x_{n,H_2}\|\left(\Gamma_1 - \Gamma_2\right)
				$}
		\end{equation*}
		where $\Gamma_1 = m_{n-1}\lvert x_{n-1,H_2}\rvert\ + m_n P_0\lvert \frac{1}{x_n}\rvert\ + G_n \lvert u_1\rvert\ + \lvert \zeta_{n,H_2}\rvert\ $, $\Gamma_2 = m_{n}\left(1+\rho_n\right) + m_{n-1} \lvert \frac{x_n x_{n-1, H_2}}{x_{n-1}}\rvert\ $, $\tilde x_{n,H_2}=x_{n,H_2} - x_{nd,H_2}$ and $\triangle_{\varphi}=min(\beta_1, \beta_2)$ is a finite positive constant.
	\end{Remark}
	
	\begin{Lemma}
		Suppose $f_1\left(\bar b (x), x\right) = \frac{\partial F_1}{\partial \bar b}$ and $f_2\left(\bar c (x), x\right) = \frac{\partial F_1}{\partial \bar c}$. If the conditions \eqref{Equilibrium_Assignment_Condition} and \eqref{Lyapunov_Stability_Condition} are satisfied by $f_1\left(\bar b (x), x\right)$ and $f_2\left(\bar c (x), x\right)$, then one of the possible selections of these functions are given as follows:
		\begin{equation}
			f_1\left(\bar b (x), x\right)=-\frac{\zeta_{1,H_2}\left[1+ln(b(x_d))\right]}{a_{11}a_{17}} - \frac{x_{smd,H_2k_{sm,H_2}}}{a_{11}},  \label{function_f_1_selection}
		\end{equation}
		\begin{equation}
			f_2\left(\bar c (x), x\right)=\frac{\zeta_{q,H_2}\left[1+ln(c(x_d))\right]}{a_{31}a_{33}} + \frac{x_{qd,H_2k_{q,H_2}}}{a_{31}}.  \label{function_f_2_selection}
		\end{equation}
	\end{Lemma}
	\begin{Proof}
		See Appendix C.
	\end{Proof}
	We propose an extended state-feedback control law based on \eqref{eq:16} and by injecting some additional damping that incorporates the collocated output $y$ as feedback. 
	\begin{equation}
		\begin{split}
			u=&G^{-1}\left[\left(J_d - R_d\right)\Omega (x) + \left(J_a - R_a\right)\nabla H (x) - \zeta \right] \\&- R_{ai}\left(e + y_d\right)
		\end{split}
		\label{extended_control_action}
	\end{equation}
	where $R_{ai}$ is a diagonal matrix of dimension 2x2 with diagonal entries are $k_{n1}$ and $k_{sm1}$.
	\begin{Remark}
		The matrix $R_{ai}$ in \eqref{extended_control_action} is having positive semi-definiteness property, which allows modification in the dissipation parameters of the closed-loop system for the measurable collocated outputs.
	\end{Remark}
	\begin{Remark}
		The denominators ($\Omega (x)$) of the designed controller actions in \eqref{extended_control_action} are state-dependents, therefore, to define the acceptable controller configuration, the constraints $x_{sm}(0)>x_1(0)$, $\left(\mu_1\rho_1+\mu_1m_{ej}\right)x_{1}(0) + \mu_2 m_1 x_{2}(0) > 0$, $x_{(q-1)}(0) > x_q(0)$, $\mu_q m_q x_{n}(0) - \left(\mu_q m_q+\rho_q\right)x_{(n-1)}(0) > 0$, and $x_{n,H_2}(0) - x_{nd,H_2}(0) > 0$ are to be satisfied to avoid the singularity conditions.
	\end{Remark}
	Based on the control designed in \eqref{extended_control_action}, the dynamics of tracking error $e(t)$ can be obtained as
	\begin{equation}
		\label{Dynamics_tracking_error}
		\dot e=C\left(J_d - R_d\right)\nabla H_d(x) - CGR_{ai}\left(e + y_d\right) -\dot y_d.
	\end{equation}
	
	\subsection{Stability Analysis} 
\begin{figure}[!t]\centering
		\includegraphics[width=8cm, height=5cm]{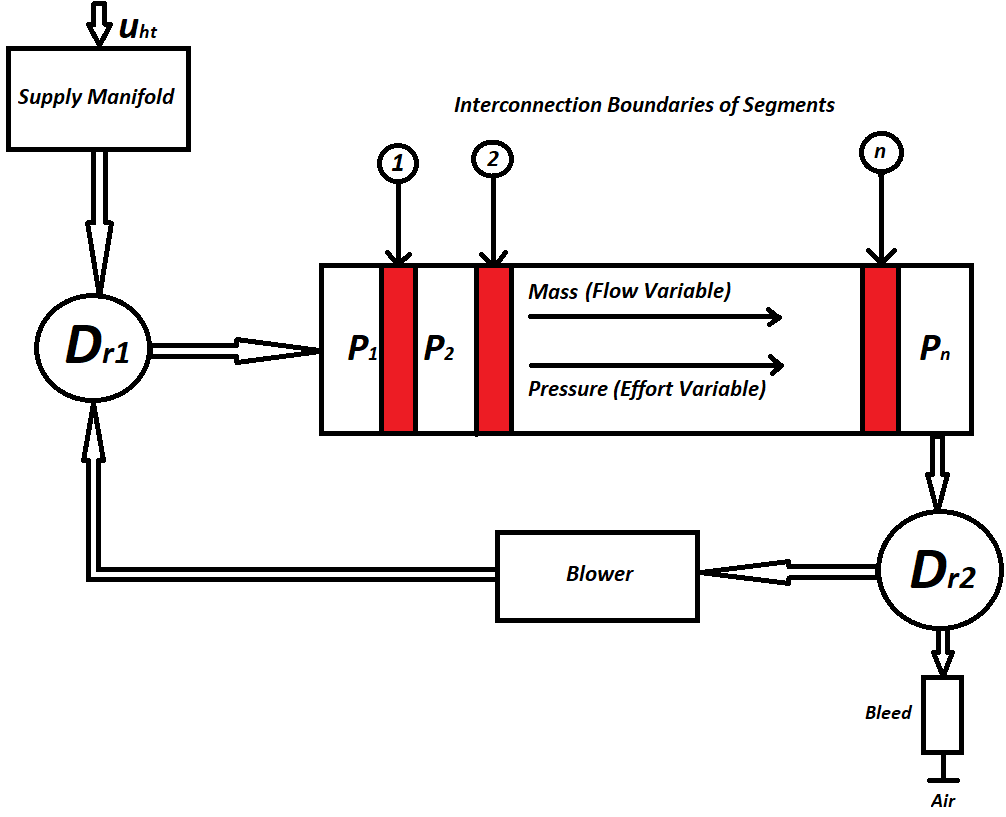}
		\caption{Geometrical PH Structure of FDS}
		\label{PEMFC_Geometrical_Structure_FDS}
	\end{figure}

	Before going into the explicit analysis of stability for the state-feedback closed-loop system, we would like to focus on the passivity of the segments as well as the structure preservation of the PCH model of segmented FDS, which in turn proves that the segments are feasible for the interconnection. The port-based geometrical model with nodes ($1, 2, ... , n$) and Dirac structures ($D_{r1}$ and $D_{r2}$) is developed for the FDS of PEM fuel cells, as shown in Fig. \ref{PEMFC_Geometrical_Structure_FDS}, which shows the structure preservation and describe the flow and effort pH variables. As the mass flow through the segments is considered as the main phenomenon, the flow and effort variables will be described as the mass of the ideal gases and spatially distributed pressures in the FDS segments. 
	\begin{Assumption}
		The spatially distributed pressure in segments are of decreasing magnitudes from the first to the $n^{th}$ segment due to resistance to the flow for the pressures, implying that they follow the relationship $x_1$ $> ... >$ $x_q$ $> ... >$ $x_n$. Furthermore, the incoming mass flow rate at each node is to be consistent with the outgoing mass flow rate. In addition, the output $y_{si}$ of each segment is contained in $L_2$ - space such that $\int_{0}^{\infty} y_{si}^T(t)y_{si} < \infty$, with $i = 1,2,...,n$ indicates segment number. 
	\end{Assumption}
	Furthermore, the passivity of the segments is proved in the following lemma with the help of Assumption 1.
	\begin{Lemma}
		The segments of FDS in (16) are passive for consistent flow of mass and to control the spatially distributed pressures. Hence, there exist a positive definite Hamiltonian function in each segment of the FDS to preserve the passivity of each interconnected node (segments), and satisfies the following conditions:
		\begin{enumerate}
			\item \label{item:mumble} $\dot H = \nabla H (x) \dot x \leq u^Ty \leq \triangle_H$ 
			\item $u^T_{s1}$$y_{s1}$ $> ... >$ $u_{sq}^T$$ y_{sq}$ $> ... >$ $u_{sn}^T$$ y_{sn}$ 
		\end{enumerate}
		where $\triangle_H$ is a small finite positive constant, and $u_{si}$ are the input to the segments. 
	\end{Lemma}
	\begin{Proof}
		In reference to \eqref{energy_open-loop}, we consider a positive definite Hamiltonian function for the each segment as follow:
		\begin{equation}
			\label{segments_Hamiltonian}
			H_{si} = \frac{1}{2}k_{i,H_2}x^2_{i,H_2} + \frac{1}{2}k_ix^2_i .
		\end{equation}
		After some rigorous calculations and utilizing Assumption 1 for $y_{si}$ to be in $L2$- space, it can be proved from \eqref{segments_Hamiltonian} that
		\begin{equation}
			\dot H_{si} = \nabla H_{si} (x)\dot x  \leq u^T_{si}y_{si}.
			\label{passivity_segments_1}
		\end{equation}
		The derivative of the total Hamiltonian can be calculated by summing up all the individual derivatives of the Hamiltonians' of each section as follows:
		\begin{equation}
			\dot H = \sum_{i=1}^{n}\nabla H_{si} (x)\dot x_{si} +u^T_{sm}y_{sm} \leq \sum_{i=1}^{n} u^T_{si}y_{si} + u^T_{sm}y_{sm} \leq u^Ty
			\label{passivity_segments_2}
		\end{equation}
		where $i=1, ... , q, ... , n$. Furthermore define $u^T_{si}y_{si} =\epsilon_{si}$ and $u^T_{sm}y_{sm} =\epsilon_{sm}$, and using $u^Ty = u^T_{s1}y_{s1} + ... + u^T_{sq}y_{sq} + ... + u^T_{sn}y_{sn} + u^T_{sm}y_{sm} = \sum_{i=1}^{n} u^T_{si}y_{si} + u^T_{sm}y_{sm}$, it can be written that $u^Ty = \epsilon_{s1} + ... + \epsilon_{sq} + ... + \epsilon_{sn} + \epsilon_{sm} = \triangle_H$. Therefore, $\dot H \leq \triangle_H$,
		this proves the passivity of each segment and consequently, the feasibility of segments interconnection. In addition, the pressures are decreasing in magnitudes as per Assumption 1, the inequality relation $u^T_{s1}$$y_{s1}$ $> ... >$ $u_{sq}^T$$ y_{sq}$ $> ... >$ $u_{sn}^T$$ y_{sn}$ will be followed.
	\end{Proof}
	From \eqref{energy_open-loop} and \eqref{passivity_segments_2}, it can be calculated that $\dot H=\nabla H (x) \dot x \leq u^Ty \leq \triangle_H$.
	Consequently, the Hamiltonian function $H$ will be a passive and bounded in a small domain of interest. The open-loop system's passivity is maintained in the presence of exogenous inputs, as well as by injecting additional damping and revamping the dissipating elements.
Since $H_d = H + H_a$, the gradient $\nabla H_d = \nabla H + \Omega (x)$ and the Hessian matrix  $\nabla^2 H_d = \nabla^2 H + \nabla \Omega (x)$ can be calculated. The gradient $\nabla H_d (x_d)=0$ and also the Hessian matrix $\nabla^2 H_d(x)$ is positive definite at $x=x_d$.
Consequently, the desired equilibrium points $x_d =$ $arg$ $min$ $(H_d(x))$ are stable and $H_d(x)$ qualify for a suitable Lyapunov candidate. 
The asymptotic stability of $x_d$ is achieved along the trajectories of \eqref{state-feedback_closed-loop}. The time derivative of Lyapunov function $H_d$ can be calculated as follows [16]: 
\begin{equation}
	\dot H_d=\nabla H_d(x) \dot x = -\nabla^T H_d(x) R_d(x)\nabla^T H_d(x)\leqslant 0  \label{Hd_negative_semidefiniteness}	
\end{equation}
which can be proved to be negative semi-definite in a compact set $\left\{x \in \mathbb R^n \mid H_d(x) \le l_d\right\}$ with $l_d$ is a small positive constant using La Salle's invariance principle and hence, the pH system \eqref{segmented_model_pH_framework} along with state-feedback control law \eqref{extended_control_action} is asymptotically stable.
\begin{Theorem}
	The controller gain $R_{ai}$ is tuned properly and there exist a Lyapunov function $V_d(x)=\frac{1}{2}e^Te + H_d (x)$ such that the error dynamics system \eqref{Dynamics_tracking_error} is bounded and converges to the desired trajectory $x=x_d$. Consequently, the pH system \eqref{segmented_model_pH_framework} along with state-feedback control law \eqref{extended_control_action} is bounded in a small DOA.
\end{Theorem}
\begin{Proof}
	See Appendix D.
\end{Proof}
\section{Output Feedback Control Design}
\subsection{Observer Design}
In practical applications, some of the states are not measurable at the output and only the partial pressures $\mathcal{P}_{an}$ and $\mathcal{P}_{sm}$ are known at the output. Therefore, it is necessary to estimate other states with the help of an observer or estimator. It is assumed that the disturbance signals are known, and the method to design the observer in this paper has been followed from \cite{RAKHTALA2014203}, and \cite{Utkin} supports finite time convergence and attenuation of chattering phenomenon. The observer for $x(t)$ in (12) is designed as follows:
\begin{equation}\label{high_order_smo}
	\begin{split}
		&\dot{\hat{x}}=\pi(\hat x)+G(x_{n,H_2}, y)u+\zeta + l_{ob}(\hat x)\nu_{ob,in}\\&
		\hat y=C\hat x  
	\end{split}	
\end{equation}
where $\pi(x)=\left[J(x)-R(x)\right]\frac{\partial H}{\partial x}(x)$. Here, $\nu_{ob,in}\in \mathbb{R}^2$  is input to the observer system which will be designed with SMC approach and the gain matrix $l_{ob}(\hat x)$ is
\begin{equation}
	\resizebox{0.8\hsize}{!}{$
		\l_{ob}(\hat x)=
		{\begin{bmatrix}
				\left(\frac{\partial Cx}{\partial x}\right)^T \\
				\left(\frac{\partial C\dot x}{\partial x}\right)^T \\
				\left(\frac{\partial Cx^{(2)}}{\partial x}\right)^T \\
				\left(\frac{\partial Cx^{(3)}}{\partial x}\right)^T	
		\end{bmatrix} }^{-1}_{x=\hat{x}}
		{\begin{bmatrix}
				0 & 0 & 0 & 0 & 0 & 0 & 1 & 0\\
				0 & 0 & 0 & 0 & 0 & 0 & 0 & 1\\		
			\end{bmatrix}^T }.\label{observer_gain} $}
\end{equation}
For the known system outputs and disturbances similar to \cite{doi:10.1080/00207170802590531}, the error in states observation can be written as follows:
\begin{equation}
	\dot{\tilde{x}} = \pi(x) - \pi(\hat{x}) - l_{ob}(\hat{x})\nu_{ob,in}  \label{observational_error}
\end{equation}
where $ \Upsilon \triangleq \begin{bmatrix} \tilde{y}^T & \dot{\tilde{y}}^T & \cdots & (\tilde{y}^{(n-1)})^T \end{bmatrix}^T $, $ \tilde{y}(t) \triangleq y - \hat{y}$, $ \tilde{x}(t) \triangleq x - \hat{x} $,  and $ \mathcal{D_{\alpha}} \subset \mathbb{R}^{8} $ is a compact domain. The implications $\Upsilon=0$ $\Leftrightarrow$ $\tilde{x}=0$ holds true for the observer defined in \eqref{high_order_smo}, if the matrix $l_{ob}(\hat{x})$ is non-singular in $\mathcal{D_{\alpha}}$. 
\begin{Remark}
	By using the Monte Carlo approach as discussed in \cite{RAKHTALA2014203}, the random numbers can be generated for $x\in\left[0, 2e7\right]$ to satisfy the condition $0<x_n,  x_{sm}<<2e7$ and $l_{ob}(\hat x)>0$. Therefore, it can be proved that the observer gain matrix $l_{ob}(\hat x)$ is non-singular for $\forall x \in \mathcal{D_{\alpha}}$.
\end{Remark}
According to \cite{Levant}, $\nu_{ob,in}$ to the observer system is designed as follows \cite{https://doi.org/10.1002/asjc.1827}:
\begin{equation}
	\begin{split}
		\nu_{ob,in} =& - \alpha_i \{e_3 + 3(e_2^6 + e_1^4 + |e_0|^3)^{\frac{1}{12}} sign[e_2 \\
		& + (e_1^4 + |e_0|^3)^{\frac{1}{6}} sign(e_1 + 0.5|e_0|^{\frac{3}{4}} sign(e_0) ) ] \} 
	\end{split}
	\label{observer_input_design}
\end{equation}
with $\alpha_i$ is a positive constant. The sliding differentiator of third-order with estimates $e_0(t)$, $e_1(t)$, $e_2(t)$, and $e_3(t)$ for $j=0,1,2,3$ of $j^{th}$ derivatives  can predict the $j^{th}$ time derivative of $\tilde{y}_i(t)$ to make its zero convergence in a known bounded time. 
Therefore, it can be said that there exist a Lyapunov function that will make the observer system asymptotically stable. Define the Lyapunov function as $V_{ob} = \frac{1}{2}\tilde{x}^T\tilde{x}$ with $V_{ob}(x)\in\mathbb{R}$. The derivative $\dot V_{ob}$ can be obtained as 
\begin{equation}
	\label{dot_Vob}
	\dot V_{ob} = \tilde{x}^T\left(\pi(x) - \pi(\hat{x}) - l_{ob}(\hat{x})\nu_{ob,in}\right)
\end{equation}
where \eqref{observational_error} is utilized.
The function $l_{ob}(\tilde x)\nu_{ob,in}$ is bounded to make the convergence to a small DOA as given in the expression below \cite{PANTELEY1998131}:
\begin{equation}
	\|l_{ob}(\hat x)\nu_{ob,in}\|<k_l\|\tilde x\|   \label{eq:92}
\end{equation}
where $k_{l}>0$. For the function $\pi(x)$ to be Lipschitz, the inequality $\|\pi(x) - \pi(\hat{x})\| \leq \rho_1 \|\tilde x\|$ can be obtained, where $\rho_1$ is a positive constant. Hence, the expression in \eqref{dot_Vob} can be written as follows:
\begin{equation}
	\label{NSD_Vob_dot}
	\dot V_{ob} \leq - \left(k_l - \rho_1\right) \|\tilde x\|^2.
\end{equation}
\subsection{Output Feedback Control With Performance Recovery}
Based on the estimated states by the observer in \eqref{high_order_smo}, the output feedback control can be designed according to the state feedback control law proposed in \eqref{extended_control_action} as follows:
\begin{equation}
	\begin{split}
		u=&G^{-1}\left[\left(J_d(\hat x) - R_d(\hat x)\right)\Omega (\hat x) + \left(J_a(\hat x) - R_a(\hat x)\right)\nabla H (\hat x) \right] \\&
		- G^{-1}\zeta - R_{ai}\left(e + y_d\right)
		\label{output_feedback_control_action}
	\end{split}
\end{equation}
The tracking error dynamics for the output feedback control can be written as 
\begin{equation}
\begin{split}
	\dot e &= C\left(\left(J(x)-R(x)\right)\nabla H(x) + \left(J_d(\hat{x})-R_d(\hat{x})\right)\nabla H_a(\hat{x})\right) +\\&C\left(J_a(\hat x)-R_a(\hat x)\right)\nabla H(\hat{x}) - CGR_{ai}\left(e + y_d\right) -\dot y_d.
	\label{tr_error_dynamics_closed-loop}
\end{split}
\end{equation}	
To determine the output feedback closed-loop system stability, a Lyapunov candidate function $V(x)\in\mathbb{R}$ is defined as
\begin{equation}
V(x)=V_{ob}(\tilde x)+\frac{1}{2}e^Te. \label{Lyapunov_function_assumption}
\end{equation}
The second function is defined to make the convergence of tracking error to an arbitrary small domain of interest. The derivative of \eqref{Lyapunov_function_assumption} can be obtained as follows:
\begin{equation}
\dot V(x)= \dot V_{ob}(x)+e^T\dot e.  \label{derivative_Lyapunov}
\end{equation}

\begin{Theorem}
Suppose, the functions $\pi_a(x)$ and $\pi_d(x)$ are Lipschitz such that
\begin{equation}
	\label{Condition_1}
	\left\|C\left(\pi_a(x) - \pi_a(\hat x)\right)\right\| \leq \delta_a \|\tilde x\|,
\end{equation}
\begin{equation}
	\label{Condition_2}
	\left\|C\left(\pi_d(x) - \pi_d(\hat x)\right)\right\| \leq \delta_d \|\tilde x\|,
\end{equation}
where the functions  $\pi_a(x)\triangleq\left(J_a(x) - R_a(x)\right)\frac{\partial H}{\partial x}(x)$ and $\pi_d(x)\triangleq\left(J_d(x) - R_d(x)\right)\frac{\partial H_a}{\partial x}(x)$. Also, $\delta_a$ and $\delta_d$ are small positive constants.
Then, the following inequality is obtained from \eqref{derivative_Lyapunov}:
\begin{equation}	\label{derivative_condition}
	\begin{split}
		\dot V(x) &\leq -\left(\sqrt{k_l - \rho_1}\|\tilde x\| - \sqrt{\delta-L_d}\|e\|\right)^2 \\&
		- \left[\left(\left(\delta_a + \delta_d\right)+2\sqrt{\left(\delta-L_d\right)\left(k_l - \rho_1\right)}\right)\right]  \|\tilde x\|\|e\|\\&- \left(\delta \|y_d\| + \|\dot y_d\|\right) \|e\|+M\|e\|
	\end{split}
\end{equation}	
where $\delta = \|CGR_{ai}\|$ and $M \triangleq \|Cf_d(x_d)\|$.
\end{Theorem}
\begin{Proof}
See Appendix E.	
\end{Proof}

\begin{Remark} \label{Remark-theorem2-1}
In \eqref{derivative_condition}, the expressions under square root will be positive if and only if $k_l > \rho_1$ and $\delta>L_d$, for every $\rho_1>0$ and $L_d>0$, respectively. Furthermore, as a result of \textbf{Theorem 2}, $\dot V(x)$ converge to an arbitrary small DOA. Therefore, the pH system \eqref{segmented_model_pH_framework} with output feedback controller \eqref{output_feedback_control_action} can converge to a small DOA. Also, the constant $M$ is sufficiently small and defined in the domain $D_1 \triangleq \left\{x \in \mathbb{R}^n | 0 < M < \epsilon_M\right\}$ with $\epsilon_M$ is a small positive constant, and the domain $D_2 \triangleq \left\{y \in \mathbb{R}^m | \delta \|y_d\| + \|\dot y_d\| \le \xi_{yd}\right\}$ is defined, where $\xi_{yd}$ is sufficiently large. Then, the supply manifold pressure $x_{sm}$ and the output segment pressure $x_{n}$ converge to the desired trajectories smoothly.
\end{Remark}
\begin{Remark}
The convergence of $\dot V(x)$ is ensured in a small domain of interest by selecting/tuning the control and observer gain parameters properly. The key design parameters in this paper are the controller gain matrix $R_{ai}$, the observer gain matrix $l_{ob}$,  and $\alpha_i$. These design parameters are greater than zero and are tuned properly to make the fast and smooth convergence of the pressures to the desired values. Therefore, the domain $D_{c1} \triangleq \left\{u \in \mathbb{R}^m, x \in \mathbb{R}^n | \delta = \|CGR_{ai}\| > L_d\right\}$, with $L_d > 0$ and the domains in Remark 7 are defined to make the convergence smooth and fast enough to stabilize the closed-loop system. In addition, if the parameter $M$ is sufficiently small as defined in Remark 7, the right hand side of the inequality (46) converge to a small domain near the desired trajectory.
\end{Remark}
\begin{table}[h]\footnotesize
\caption{Regression coefficient table for blower model}
\label{Blower_Regression_Coefficient}
\begin{tabular}{m{40pt}<{\centering}m{50pt}<{\centering}|m{40pt}<{\centering}m{50pt}<{\centering}}
	\midrule
	\textbf{Parameters}  &\textbf{Values} &\textbf{Parameters} &\textbf{Values}\\
	\midrule
	$a_0$  & 2.21e-03 & $c_0$  &  0.43 \\
	\midrule
	$a_1$ & -4.64e-05 & $c_1$ & -0.68  \\
	\midrule
	$a_2$ & -5.36e-04 & $c_2$ & 0.80 \\
	\midrule
	$a_3$ & 2.70e-04 & $c_3$ & -0.43 \\
	\midrule
	$a_4$ & -3.70e-04 & $c_4$ & 0.11 \\
	\midrule
	$b_0$ & 2.44 & $c_5$ & -9.79e-03 \\
	\midrule
	$b_1$ & -1.31 & $b_2$ & 1.77 \\
	\bottomrule		    
\end{tabular}
\end{table}
\begin{table}[h]\footnotesize
\caption{Values of PEMFC variables used in the simulation}
\label{Model_Design_Variables}
\begin{tabular}{m{40pt}<{\centering}m{50pt}<{\centering}|m{40pt}<{\centering}m{50pt}<{\centering}}
	\midrule
	\textbf{Physical Variables}  &\textbf{Values with Units} &\textbf{Physical Variables} &\textbf{Values with Units}\\
	\midrule
	$\alpha$  & 0.01 N-m & $\mathcal P_{sat}$  &  1.762e04 Pa \\
	\midrule
	$A_{ai}$ & 8.04e-06 $m^2$ & $\mathcal P_0$ & 1.01e05 Pa  \\
	\midrule
	$A_{bd}$ & 7.24e-05 $m^2$ & $V_a$ & 1.1e-04 $m^3$ \\
	\midrule
	$A_{or}$ & 7.24e-06 $m^2$ & $V_c$ & 1.9e-04 $m^3$ \\
	\midrule
	$d_{bl}$ & 0.2286 m & $V_{sm}$ & 1.608e-05 $m^3$ \\
	\midrule
	$F$ & 96485 c/s & $T_a$ & 298 K \\
	\midrule
	$k_{cr}$ & 7.455e-12 mol/(Pa.s) & $T_{sm}$ & 298 K \\
	\midrule
	$\dot{\eta}_{ht,m}$ & 2.5e-03 mol/s & $t_1, t_2, t_3$ & 1/3 \\
	\midrule
	$N_{fc}$ & 25  & - & - \\
	\bottomrule		    
\end{tabular}
\end{table}

\section{Simulation Results and Discussions}
In this work, a nonlinear segmented model with generalized structure is described with n segments of the anode flow channel. For two segments model ($n=2$), we do not need the observer to estimate the internal states. For at least three segments ($n=3$), an observer-based control needs to be designed. The choice of number of segments depends on the model complexity, feasibility of the control algorithm, and computational time. Moreover, the paper is not dealing with the  comparisons between the lump sum ODEs model of the segmented PEM fuel cell and PDEs model. If their is more number of segments then the system can be modeled more accurately by approximating PDEs model by ODEs model. Regarding to the system properties, the number of segments can be selected in such a way that there is reasonable distribution of the variables such as pressure, temperature, and current density which can be verified from the previous research work in \cite{CHEN202048}. Therefore taking care of these aspects, in this work, a three segments complex model is used in this paper for the control design and simulations.

The compressor regression coefficients of the model ($a_0, a_1,..., a_4$), ($c_0, c_1,..., c_5$), and ($b_0, b_1, b_2$) are provided in Table \ref{Blower_Regression_Coefficient} \cite{Pukrushpan_Thesis}. The physical parameters for simulation are selected based on the fuel cell stack manufacturer as shown in Table \ref{Model_Design_Variables}. Here, we have taken $q=2, \alpha_1=0.1, \alpha_2=1, k_{n1}G_{sm,ht}=-1e^{-3}$, $k_{sm1}G_{sm,ht}=-1e^{+4}$, $k_1 =1$, $k_{1,H_2} =1$, $k_{2,H_2} =1$, $k_2 =1$, $k_{3,H_2} =1$, and $k_3 =1$ for the simulation purpose. 

\begin{figure}[!t]\centering
\includegraphics[width=8cm, height=3cm]{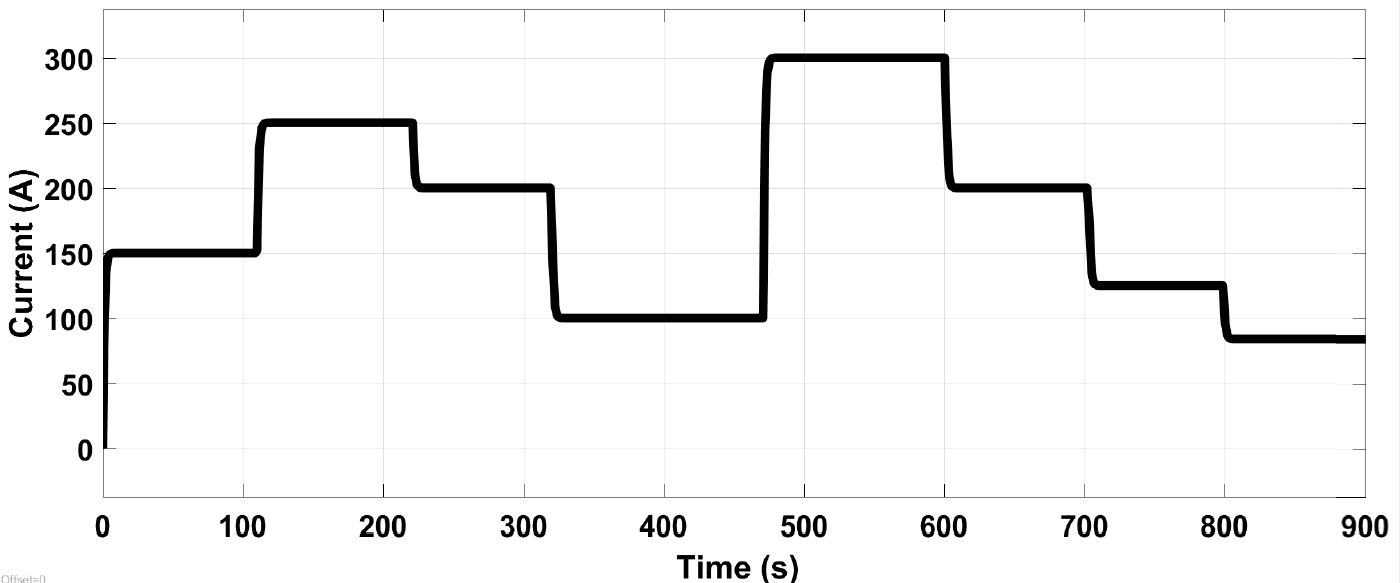}
\caption{Stack Current}
\label{fig:Stack_Current}
\end{figure}

\begin{figure}[!t]\centering
\includegraphics[width=8cm]{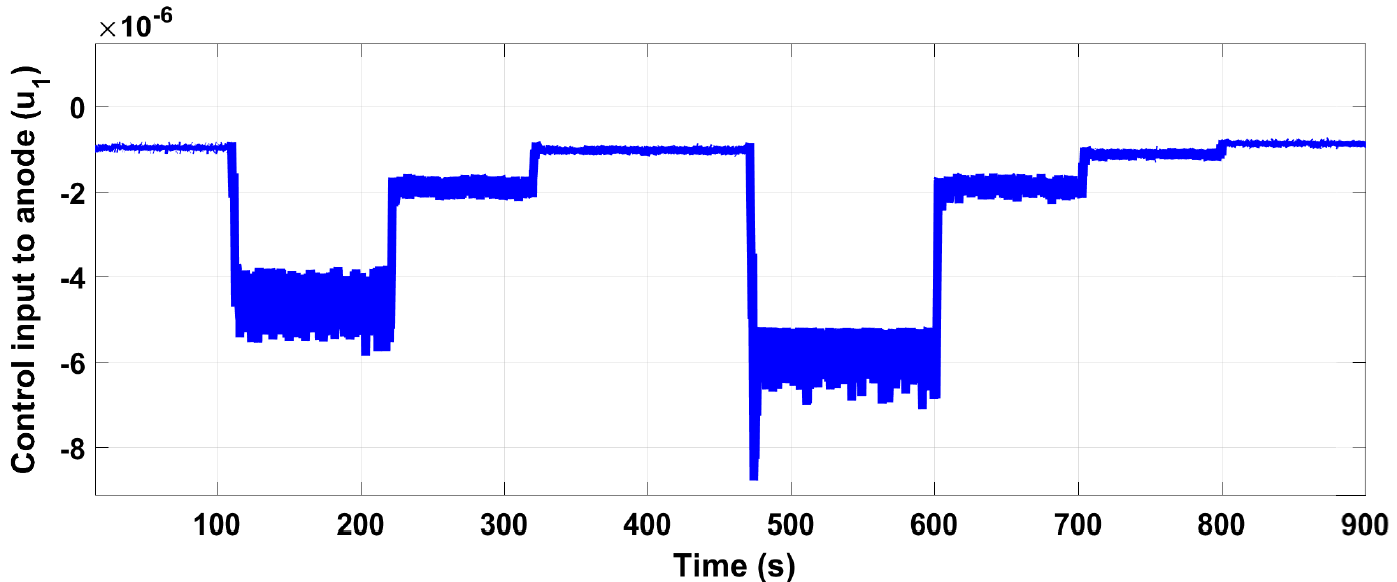}
\caption{Control input $u_1$ to the anode}
\label{fig:input1}
\end{figure}

\begin{figure}[!t]\centering
\includegraphics[width=8cm]{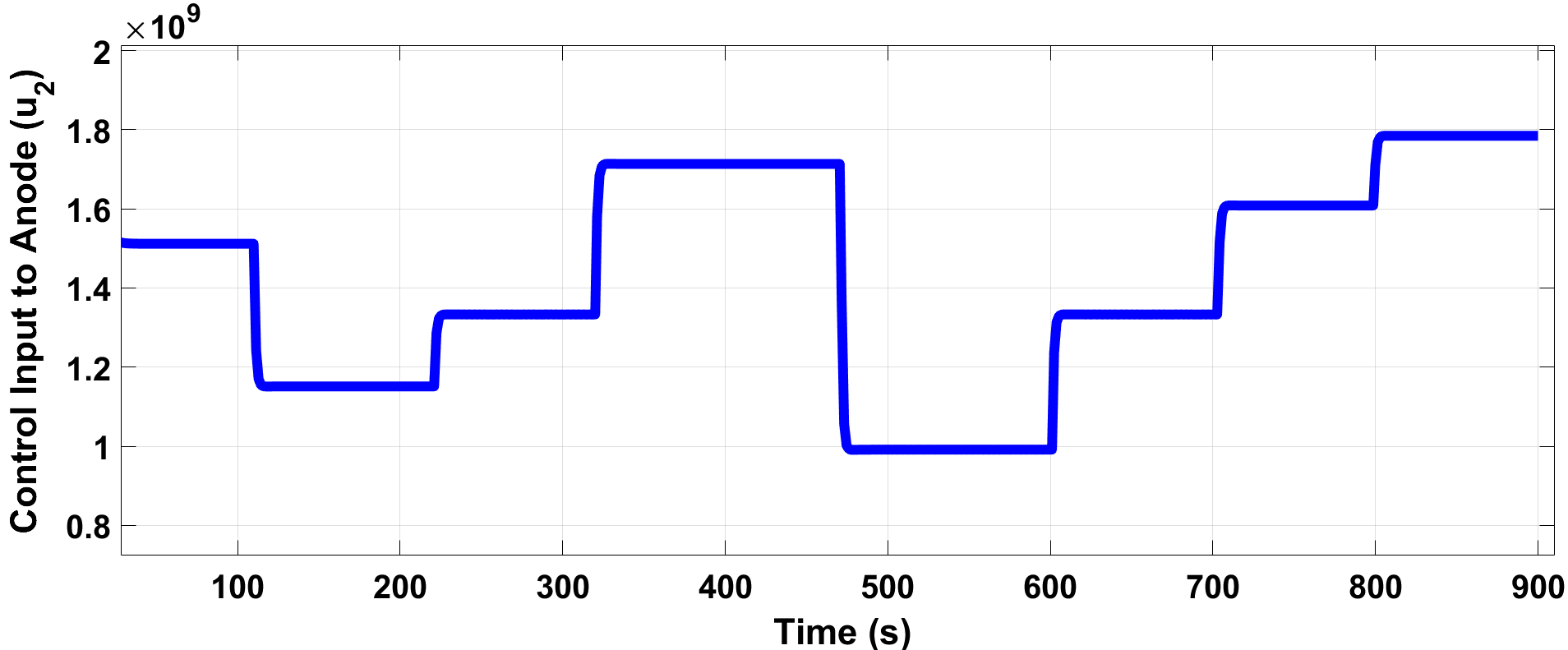}
\caption{Control input $u_2$ to the anode}
\label{fig:input2}
\end{figure}

The stack current, the controller action $u_1$ or $u_{ht}$, and the blower control action $u_2$ or $u_{bl}$ are depicted by \figurename{~\ref{fig:Stack_Current}}, \figurename{~\ref{fig:input1}}, and \figurename{~\ref{fig:input2}}. The control action in \figurename{~\ref{fig:input1}} depicts the adequate hydrogen supply with required concentration in the quickest time from the hydrogen source tank. Furthermore, during time interval shown in the figure, it needs to be regulated slowly to balance the hydrogen supply. However, the blower control action is shown in the \figurename{~\ref{fig:input2}}, it can be predicted that the hydrogen circulation and supply manifold hydrogen pressure are balanced, and fast enough to fulfill the $H_2$ requirement to PEM fuel cells. 
\begin{figure}[!t]\centering
\includegraphics[width=8cm]{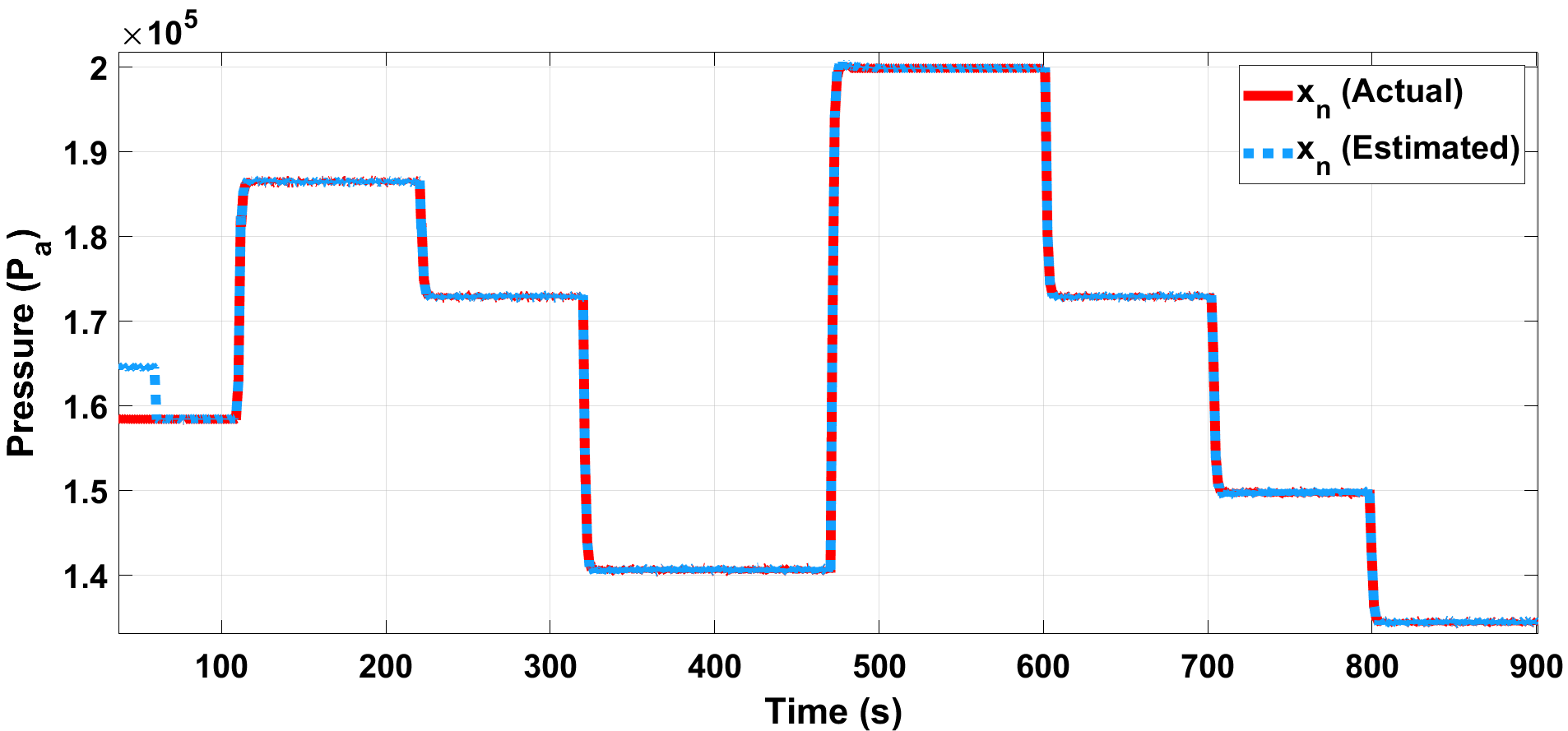}
\caption{Estimate of output pressure of the $n^{th}$ segment (anode outlet pressure)}
\label{estimate_nth}
\end{figure}
The estimate of $n^{th}$ segment pressure and the error are shown in \figurename{~\ref{estimate_nth}} and \figurename{~\ref{estimation_error_nth}}, it can be observed that the actual and measured pressures follow each other closely. As the stack current changes from 150 A to 250 A, the pressures changes accordingly.

Furthermore, \figurename{~\ref{estimate_sm}} and  \figurename{~\ref{estimation_error_sm}} show the estimate of supply manifold pressure and the estimation error, it is clear that the supply manifold pressure needs to be regulated slowly as the actual and measured values follow each other with some transitions at regular intervals. However, as compared to the pressures of $n^{th}$ segment and the supply manifold which are of the order of $10^5$ Pa, the error between actual and measured pressures are very low as shown in \figurename{~\ref{estimation_error_nth}} and \figurename{~\ref{estimation_error_sm}}. It is depicted that the maximum error in the $n^{th}$ segment is less than 350 Pa, however, in the supply manifold, it is less than 300 Pa. Hence, these errors can be considered as negligible as compared to the actual and measured pressures.

The tracking performance of outputs $x_n$ and $x_{sm}$ are shown in \figurename{~\ref{Tracking_xn}} and \figurename{~\ref{Tracking_sm}}. The tracking of anode output pressure $x_n$ is very fast but there exist some errors when current changes during the time intervals $0$ to $100$ secs, $320$ to $470$ secs, and $700$ to $900$ secs. However, in case of supply manifold pressure $x_{sm}$ the tracking is quick and accurate with minimum error.
\begin{figure}[!t]\centering
\includegraphics[width=8cm]{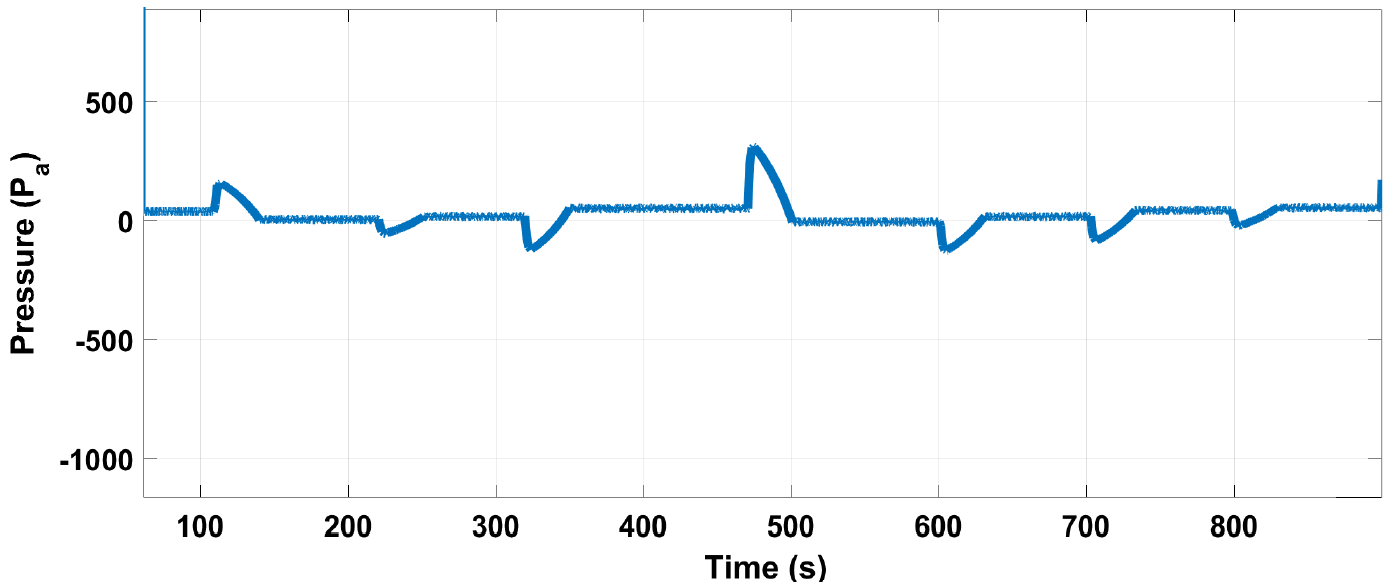}
\caption{Error between actual and estimated output pressures of the $n^{th}$ segment}
\label{estimation_error_nth}
\end{figure}
\begin{figure}[!t]\centering
\includegraphics[width=8cm]{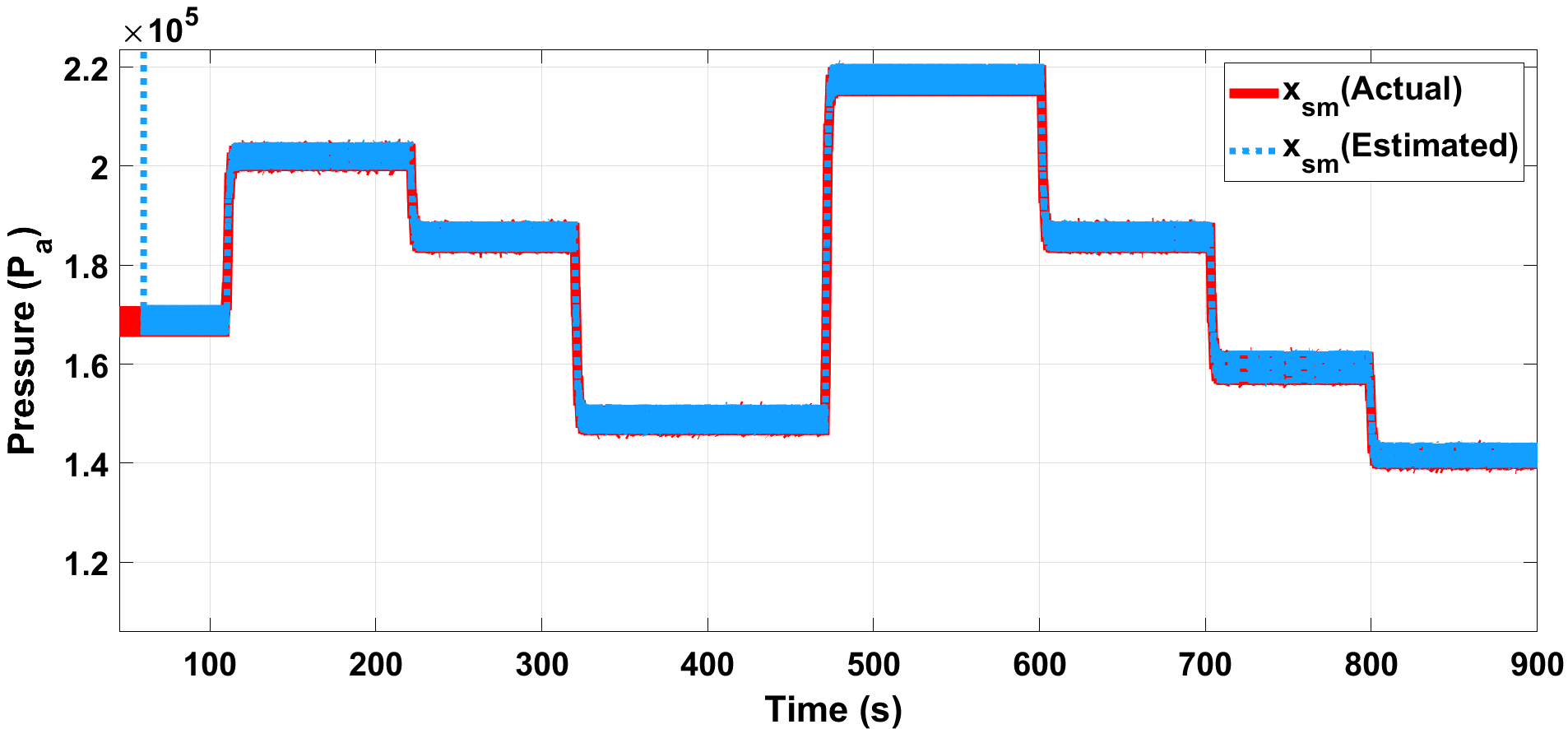}
\caption{Estimate of supply manifold pressure $P_{sm}$}
\label{estimate_sm}
\end{figure}
\begin{figure}[!t]\centering
\includegraphics[width=8cm]{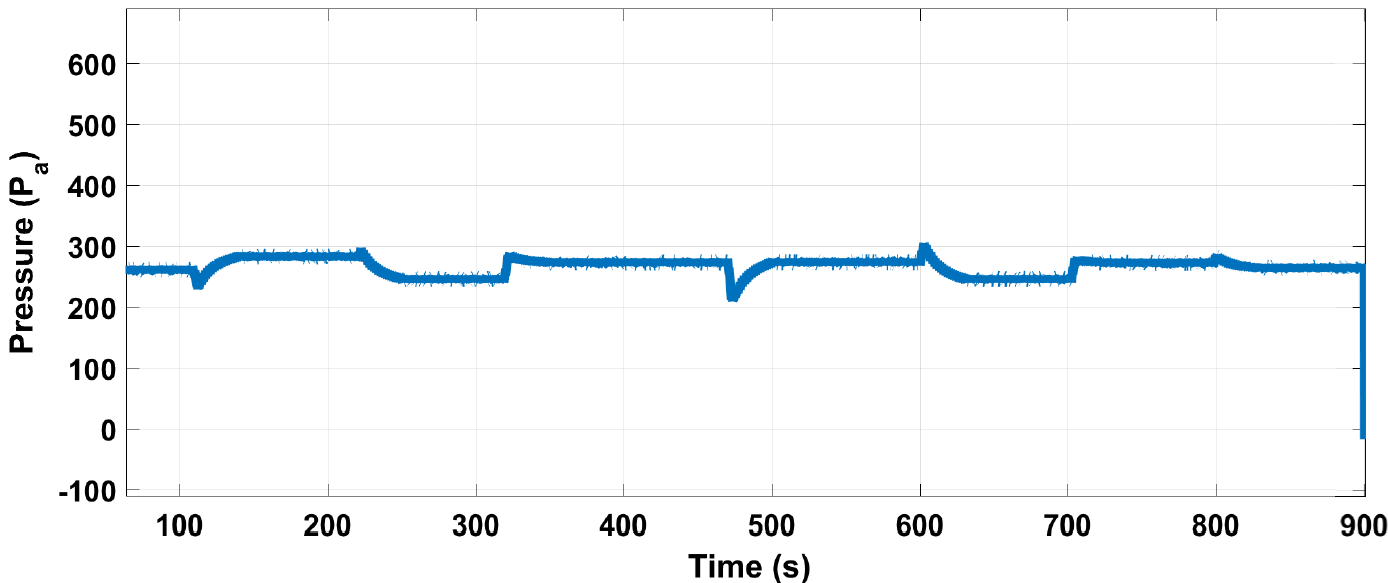}
\caption{Error between actual and estimated supply manifold pressures $P_{sm} (Pa)$}
\label{estimation_error_sm}
\end{figure}
\begin{figure}[!t]\centering
\includegraphics[width=8cm]{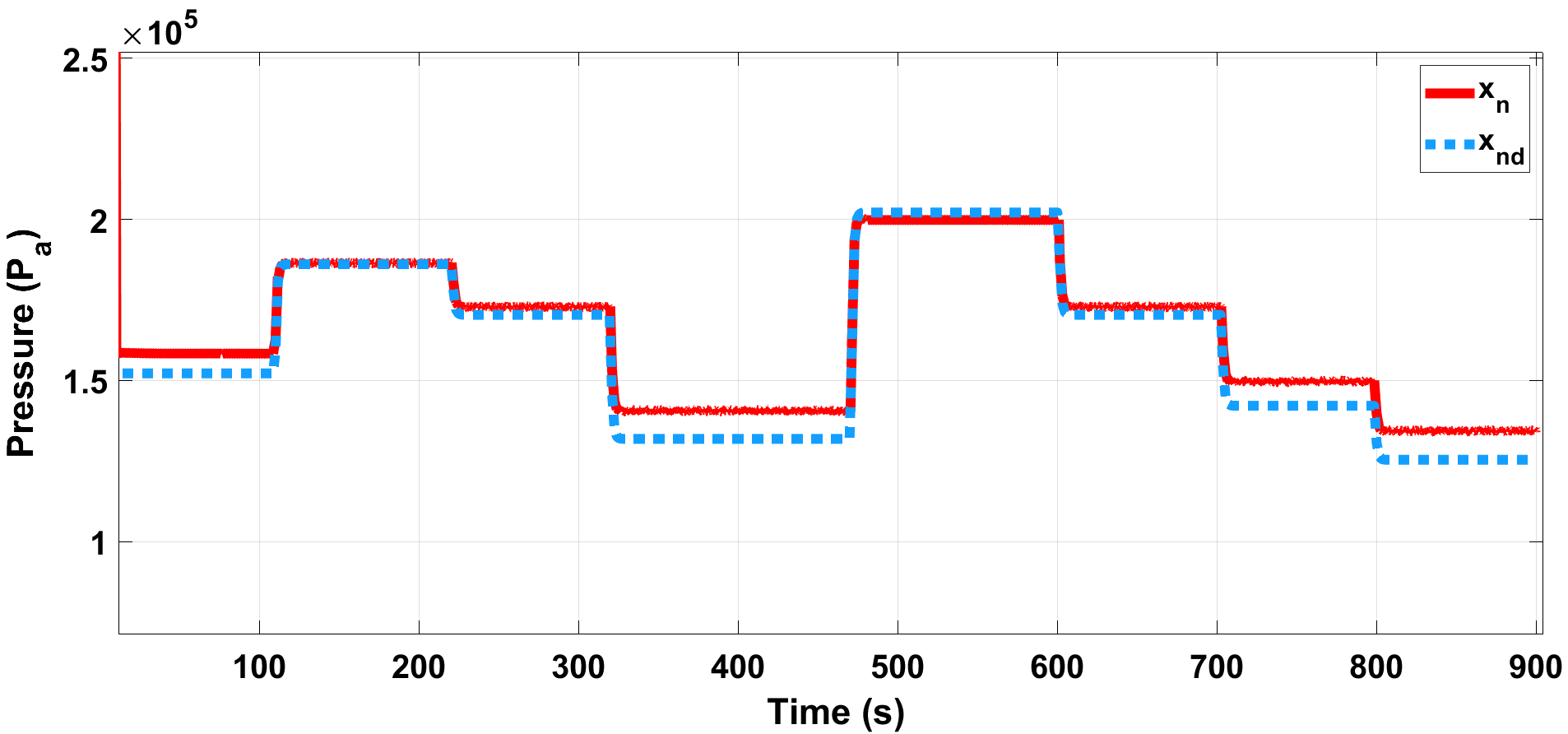}
\caption{Tracking of $n^{th}$ segment pressure $P_{an} (Pa)$}
\label{Tracking_xn}
\end{figure}

Previous research \cite{CHEN202048} addressed the reasonable distribution of physical system parameters/properties such as pressures, temperature, and current density while testing the segmented model accuracy. But, the work does not investigate the feasibility of control algorithm; instead, the work focuses on scalability testing of the model from $3 \times 3$ to $6\times 6$ segments. However, in the present work, the main focus is the feasibility of the control algorithm in port-Hamiltonian framework, which has not been examined previously for the segmented PEM fuel cell system.
\begin{figure}[!t]\centering
\includegraphics[width=8cm, height=3cm]{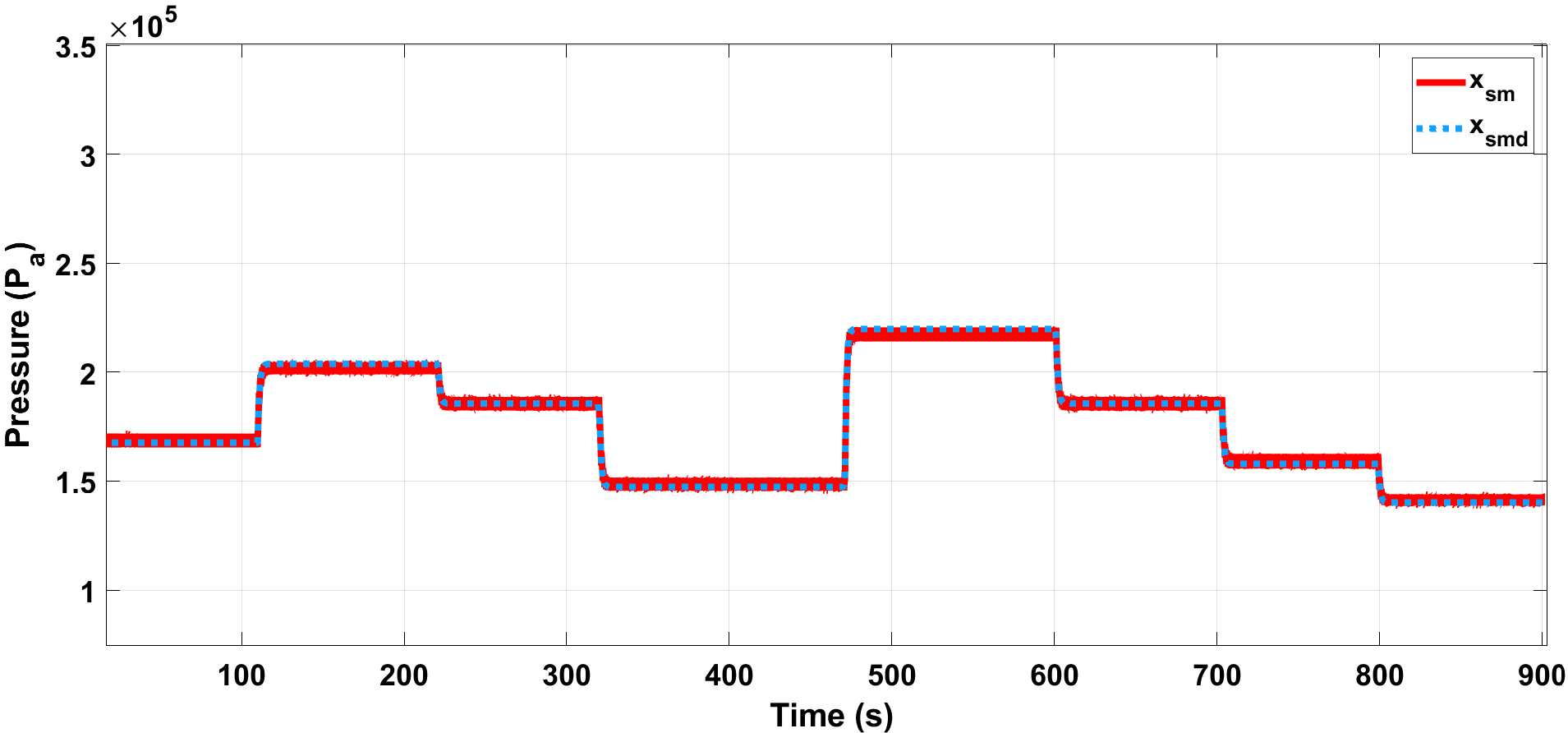}
\caption{Tracking of supply manifold pressure $P_{sm} (Pa)$}
\label{Tracking_sm}
\end{figure}
\section{Conclusions}
In this paper, an IDA-PBC based controller is proposed to control the spatially distributed pressures while maintaining the required flow rate and concentration of hydrogen. A nonlinear MIMO model has been developed in pH framework by considering spatial distribution of parameters. The proposed modeling concept minimizes the modeling error and ensured the passivity of each segment for the feasible interconnection to achieve the control objectives. Besides, the designed sliding mode observer is estimating the unmeasurable pressures at the output with less estimation error. The proposed IDA-PBC-based controller structure has recovered the desired performance of the state feedback controller. Furthermore, the stability of closed-loop controlled system has been analyzed explicitly in this paper.

\section*{Appendix} 

\subsection{Auxiliary Variables}
The disturbances are as defined below:
\begin{equation*}
\resizebox{1\hsize}{!}{$
	\zeta_{1,H_2} \triangleq -\left(\frac{RTN_{fc}}{2F}\right)\left(\frac{I_1}{V_{a1}}\right), \zeta_{1} \triangleq -\left(\frac{RTN_{fc}}{2F}\right)\left(\frac{I_1}{V_{a1}}\right)
	+ \mu_1\xi_1\left(\mathcal{P}_{c,N_2}+\mathcal{P}_v\right),\\
	$}
\end{equation*}
\begin{equation*}
\resizebox{1\hsize}{!}{$
	\zeta_{q,H_2} \triangleq -\left(\frac{RTN_{fc}}{2F}\right)\left(\frac{I_q}{V_{aq}}\right), \zeta_{q} \triangleq -\left(\frac{RTN_{fc}}{2F}\right)\left(\frac{I_q}{V_{aq}}\right)+\mu_q\xi_q\left(\mathcal{P}_{c,N_2}+\mathcal{P}_v\right),\\
	$}	
\end{equation*}
\begin{equation*}
\resizebox{1\hsize}{!}{$
	\zeta_{n,H_2} \triangleq -\left(\frac{RTN_{fc}}{2F}\right)\left(\frac{I_n}{V_{an}}\right), \zeta_{n} \triangleq -\left(\frac{RTN_{fc}}{2F}\right)\left(\frac{I_n}{V_{an}}\right)+\mu_n\xi_n\left(\mathcal{P}_{c,N_2}+\mathcal{P}_v\right).
	$}
\end{equation*}

Auxiliary variables introduced to facilitate in the formulation of the state-space model \eqref{first segment:24}-\eqref{supply_manifold:27} are as follows:
\begin{equation*}
\resizebox{1\hsize}{!}{$
	a_{11}=-\left\{\mu_1\rho_1+\mu_1 m_{1}-\mu_1 m_1\left(\frac{x_2}{x_1}\right)\right\},a_{17}=\mu_1 m_{ej}\left(1-\frac{x_1}{x_{sm}}\right),a_{24}=\mu_1 m_1, 
	$}
\end{equation*}
\begin{equation*}
\resizebox{1\hsize}{!}{$
	a_{21}=\mu_1\left(\xi_1-\rho_1\right),
	a_{22}=-\mu_1\left(m_{ej}+\xi_1+ m_1\right),a_{31}=\mu_q m_{(q-1)}\left(1-\frac{x_q}{x_{(q-1)}}\right),
	$}
\end{equation*}
\begin{equation*}
\resizebox{1\hsize}{!}{$
	a_{28}=\mu_1 m_{ej},a_{33}=-\left\{\mu_q\rho_q+\mu_q m_q\left(1-\frac{x_{(q+1)}}{x_q}\right)\right\},a_{42}=\mu_q m_{(q-1)},a_{82}=\mu_{sm}m_{ej},
	$}
\end{equation*}
\begin{equation*}
\resizebox{1\hsize}{!}{$
	a_{43}=\mu_q\left(\xi_q-\rho_q\right),a_{44}=-\mu_q\left( m_{(q-1)}+\xi_q+m_q\right),a_{46}=\mu_q m_q,a_{64}=\mu_n m_{(n-1)},
	$}
\end{equation*}
\begin{equation*}
\resizebox{1\hsize}{!}{$
	a_{53}=\mu_n m_{(n-1)}\left(1-\frac{x_n}{x_{(n-1)}}\right),a_{55}=-\mu_n\left( m_n\left(1-\frac{P_0}{x_n}\right)+\rho_n\right),a_{65}=\mu_n\left(\xi_n-\rho_n\right),
	$}	
\end{equation*}
\begin{equation*}
\resizebox{1\hsize}{!}{$
	a_{66}=-\mu_n\left( m_{(n-1)}+ m_n+\xi_n\right),a_{77}=-\mu_{sm}m_{ej}\left(1-\frac{x_1}{x_{sm}}\right),a_{88}=-\mu_{sm}m_{ej}
	$}	
\end{equation*}
with $	\mu_1 \triangleq \frac{RT_a}{V_{a1}}$,  $\mu_q \triangleq \frac{RT_a}{V_{aq}}$, $\mu_n \triangleq \frac{RT_a}{V_{an}}$, $\mu_{sm} \triangleq \frac{RT_{sm}}{V_{sm}}$, $m_{ej} \triangleq \frac{\alpha A_{ai}}{M_{ej}}$, $m_1 \triangleq \frac{\alpha A_{a1}}{M_{a1}}$, $m_q \triangleq \frac{\alpha A_{aq}}{M_{aq}}$, $m_n \triangleq \frac{\alpha A_{bd}}{M_{an}}$, $\rho_1 \triangleq {t_1 N_{fc} k_{cr,H_2}}$, $\rho_q \triangleq {t_q N_{fc} k_{cr,H_2}}$, $\rho_n \triangleq {t_n N_{fc} k_{cr,H_2}}$, $\xi_1 \triangleq {t_1 N_{fc} k_{cr,N_2}}$, $\xi_q \triangleq {t_q N_{fc} k_{cr,N_2}}$, and $\xi_n \triangleq {t_n N_{fc} k_{cr,N_2}}$.
\subsection{Interconnection and Damping Matrices}

The damping and interconnection matrices of Hamiltonian system \eqref{eq:16} is given as following:
\begin{equation*}
\resizebox{1\hsize}{!}{$
	J(x)=
	{\begin{bmatrix}
			0 & -\frac{a_{21}}{2} & -\frac{a_{31}}{2} & 0 & 0 & 0 & \frac{a_{17}}{2} & 0\\
			\frac{a_{21}}{2} & 0 & 0 & \frac{a_{24}-a_{42}}{2} & 0 & 0 & 0 & \frac{a_{28}-a_{82}}{2}\\
			\frac{a_{31}}{2} & 0 & 0 & -\frac{a_{43}}{2} & -\frac{a_{53}}{2} & 0 & 0 & 0\\
			0 & \frac{a_{42}-a_{24}}{2} & \frac{a_{43}}{2} & 0 & 0 & \frac{a_{46}-a_{64}}{2} & 0 & 0\\
			0 & 0 & \frac{a_{53}}{2} & 0 & 0 & -\frac{a_{65}}{2} & 0 & 0\\
			0 & 0 & 0 & \frac{a_{64}-a_{46}}{2} & \frac{a_{65}}{2} & 0 & 0 & 0\\
			-\frac{a_{17}}{2} & 0 & 0 & 0 & 0 & 0 & 0 & 0\\
			0 & \frac{a_{82}-a_{28}}{2} & 0 & 0 & 0 & 0 & 0 & 0
	\end{bmatrix} }, 
	$}
\end{equation*}
\begin{equation*}
\resizebox{1\hsize}{!}{$
	R(x)=	
	{\begin{bmatrix}
			-a_{11} & -\frac{a_{21}}{2} & -\frac{a_{31}}{2} & 0 & 0 & 0 & -\frac{a_{17}}{2} & 0\\
			-\frac{a_{21}}{2} & -a_{22} & 0 & -\frac{a_{24}+a_{42}}{2} & 0 & 0 & 0 & -\frac{a_{28}+a_{82}}{2}\\
			-\frac{a_{31}}{2} & 0 & -a_{33} & -\frac{a_{43}}{2} & -\frac{a_{53}}{2} & 0 & 0 & 0\\
			0 & -\frac{a_{42}+a_{24}}{2} & -\frac{a_{43}}{2} & -a_{44} & 0 & -\frac{a_{46}+a_{64}}{2} & 0 & 0\\
			0 & 0 & -\frac{a_{53}}{2} & 0 & -a_{55} & -\frac{a_{65}}{2} & 0 & 0\\
			0 & 0 & 0 & -\frac{a_{64}+a_{46}}{2} & -\frac{a_{65}}{2} & -a_{66} & 0 & 0\\
			-\frac{a_{17}}{2} & 0 & 0 & 0 & 0 & 0 & -a_{77} & 0\\
			0 & -\frac{a_{82}+a_{28}}{2} & 0 & 0 & 0 & 0 & 0 & -a_{88}
	\end{bmatrix} }. 
	$}
\end{equation*}

The assigned matrices are as follows: 

\begin{equation*}
J_a(x)=
{\begin{bmatrix}
		0 & 0 & 0 & 0 & 0 & 0 & 0 & 0\\
		0 & 0 & 0 & 0 & 0 & 0 & 0 & 0\\
		0 & 0 & 0 & 0 & 0 & 0 & 0 & 0\\
		0 & 0 & 0 & 0 & 0 & 0 & 0 & 0\\
		0 & 0 & 0 & 0 & 0 & 0 & 0 & 0\\
		0 & 0 & 0 & 0 & 0 & 0 & -a_{76} & 0\\
		0 & 0 & 0 & 0 & 0 & a_{76} & 0 & -a_{87}\\
		0 & 0 & 0 & 0 & 0 & 0 & a_{87} & 0
\end{bmatrix} }. 
\end{equation*}
\begin{equation*}
R_a(x)=	
{\begin{bmatrix}
		0 & 0 & 0 & 0 & 0 & 0 & 0 & 0\\
		0 & 0 & 0 & 0 & 0 & 0 & 0 & 0\\
		0 & 0 & 0 & 0 & 0 & 0 & 0 & 0\\
		0 & 0 & 0 & 0 & 0 & 0 & 0 & 0\\
		0 & 0 & 0 & 0 & 0 & 0 & 0 & 0\\
		0 & 0 & 0 & 0 & 0 & k_{66} & 0 & 0\\
		0 & 0 & 0 & 0 & 0 & 0 & 0 & 0\\
		0 & 0 & 0 & 0 & 0 & 0 & 0 & k_{88}
\end{bmatrix} }
\end{equation*}
with assigned damping $k_{66} = - G_n k_{n1}x_{n,d} \left(\Omega_{n} + k_n x_n\right)^{-1}$ and $k_{88} = G_{sm,ht} k_{sm1} x_{sm,d}\left(\Omega_{sm} + k_{sm}x_{sm}\right)^{-1}$. Since $J_d(x)=J(x)+J_a(x)$ and $R_d(x)=R(x)+R_a(x)$, we can calculate $J_a$ and $R_a$ with $a_{76}=\frac{a_{31}}{2}$ and $a_{87}=\frac{a_{21}}{2}$.

\subsection{Proof of Lemma 1}
The functions $F_1$ and $F_2$ are appeared in \eqref{eq:36} satisfy the conditions defined in \eqref{Equilibrium_Assignment_Condition} and \eqref{Lyapunov_Stability_Condition}, respectively. By utilizing the condition in \eqref{Equilibrium_Assignment_Condition} and the relation $\Omega (x)=\nabla H_a(x)$, the following expressions can be obtained at $x=x_d$:
\begin{equation}
\frac{\partial F_1}{\partial \bar b}\mid_{x=x_d}=-\frac{\zeta_{1,H_2}\left[1+ln(b)\right]}{a_{11}a_{17}}\mid_{x=x_d} - \frac{x_{sm,H_2k_{sm,H_2}}}{a_{11}}\mid_{x=x_d},  \label{calculation_F1}
\end{equation}
\begin{equation}
\frac{\partial F_2}{\partial \bar c}\mid_{x=x_d}=\frac{\zeta_{q,H_2}\left[1+ln(c)\right]}{a_{31}a_{33}}\mid_{x=x_d} + \frac{x_{q,H_2k_{q,H_2}}}{a_{31}}\mid_{x=x_d}.  \label{calculation_F2}
\end{equation}
After incorporating the Lyapunov stability condition \eqref{Lyapunov_Stability_Condition}, the following inequalities are obtained:
\begin{equation}
\frac{\partial^2 F_1}{\partial x^2_{sm,H_2}}(x_d) > \frac{\zeta_{1,H_2}a_{11}}{b\ a_{17}}\mid_{x=x_d} + k_{sm,H_2},  \label{double_derivative_F1}
\end{equation}
\begin{equation}
\frac{\partial^2 F_2}{\partial x^2_{q,H_2}}(x_d) > \frac{\zeta_{q,H_2}a_{31}}{c\ a_{33}}\mid_{x=x_d}+k_{q,H_2}.  \label{double_derivative_F2}
\end{equation}

For the selected $\varphi\left({x_{n,H_2}}\right)$ in \eqref{eq:36}, the minimum of the shaped energy function can be achieved if \eqref{calculation_F1}-\eqref{double_derivative_F2} are satisfied with $\frac{\partial^2 F_1}{\partial x^2_{sm,H_2}}(x_d) > 0$ and $\frac{\partial^2 F_2}{\partial x^2_{q,H_2}}(x_d) > 0$. Hence, an obvious selection of functions $f_1\left(\bar b (x), x\right)$ and $f_2\left(\bar c (x), x\right)$ in \eqref{function_f_1_selection} and \eqref{function_f_2_selection} can be obtained.

\subsection{Proof of Theorem 1}
The derivative of $V_d$ can be written as
\begin{equation}
\dot V_d(x)= e^T \dot e + \dot H_d (x).
\label{Vd_derivative}
\end{equation}
By substituting the error dynamics \eqref{Dynamics_tracking_error} in \eqref{Vd_derivative} and defining $\delta = \|CGR_{ai}\|$, the following expression can be obtained:
\begin{equation}
\resizebox{1\hsize}{!}{$
	\dot V_d \leq \|e\| \|CF_d(x)\nabla H_d(x)\| - \delta \left(\|e\|^2 + \|e\| \|y_d\|\right) - \|e\| \|\dot y_d\| + \dot H_d.
	$}
\label{derivative_Vd_inequality}
\end{equation}
Define $ f_d(x)\triangleq F_d(x)\nabla H_d(x)$ and the function $f_d(x)$ is Lipschitz such that
\begin{equation}
\|Cf_d(x)\| \leq L_d\|x - x_d\| + \|Cf_d(x_d)\| < L_d\|e\|+M
\label{upper_bound_Cf_d}
\end{equation}
where $M \triangleq \|Cf_d(x_d)\|$ and $L_d$ are small positive constants. By substituting \eqref{upper_bound_Cf_d} in \eqref{derivative_Vd_inequality} and also using the relation $\dot H_d(x) \le 0$ from \eqref{Hd_negative_semidefiniteness}, we can get
\begin{equation}\resizebox{1\hsize}{!}{$
	\dot V_d \le -\left(\delta-L_d\right) \left(\|e\| - \frac{M}{2\left(\delta-L_d\right)}\right)^2 - \left(\delta\|y_d\| + \|\dot y_d\|\right) \|e\| + \frac{M^2}{4\left(\delta-L_d\right)} 
	$}.
\end{equation}
The controller gains are tuned properly such that $\delta>L_d$, for every $L_d>0$ and M is sufficiently small. This proves the convergence of error in \eqref{Dynamics_tracking_error} and the pH system \eqref{segmented_model_pH_framework} along with state-feedback control law \eqref{extended_control_action} in a small DOA.
\subsection{Proof of Theorem 2}
Substitute \eqref{NSD_Vob_dot} and \eqref{tr_error_dynamics_closed-loop} into \eqref{derivative_Lyapunov} and using the definitions $\pi (x)$, $\pi_a (x)$, and $\pi_d (x)$, the following inequality can be obtained
\begin{equation}
\begin{split}
	&\dot V(x) \leq  - \delta \|e\|^2 + \|e^T \left[C\left(F_d(x)\nabla H_d(x)\right)\right]\| \\& - \|e^TC\left(\pi_a (x) - \pi_a (\hat x)\right)\| 
	- \|e^TC\left(\pi_d (x) - \pi_d (\hat x)\right)\| \\&- \left(\delta \|y_d\| + \|\dot y_d\| \right)\|e\| - \left(k_l - \rho_1\right) \|\tilde x\|^2 .
\end{split}
\label{theorem2_first_equation}
\end{equation}
Based on the conditions defined in  \eqref{Condition_1}, \eqref{Condition_2}, and \eqref{upper_bound_Cf_d},  the expression given in \eqref{derivative_condition} can be obtained, which completes the proof of \textit{Theorem 2}.

\bibliography{arXivbib} 

\begin{thebibliography}{10}
\providecommand{\url}[1]{#1}
\csname url@samestyle\endcsname
\providecommand{\newblock}{\relax}
\providecommand{\bibinfo}[2]{#2}
\providecommand{\BIBentrySTDinterwordspacing}{\spaceskip=0pt\relax}
\providecommand{\BIBentryALTinterwordstretchfactor}{4}
\providecommand{\BIBentryALTinterwordspacing}{\spaceskip=\fontdimen2\font plus
\BIBentryALTinterwordstretchfactor\fontdimen3\font minus
  \fontdimen4\font\relax}
\providecommand{\BIBforeignlanguage}[2]{{%
\expandafter\ifx\csname l@#1\endcsname\relax
\typeout{** WARNING: IEEEtran.bst: No hyphenation pattern has been}%
\typeout{** loaded for the language `#1'. Using the pattern for}%
\typeout{** the default language instead.}%
\else
\language=\csname l@#1\endcsname
\fi
#2}}
\providecommand{\BIBdecl}{\relax}
\BIBdecl

\bibitem{Sharma1}
S.~Sharma and S.~K. Ghoshal, ``{Hydrogen the future transportation fuel: From
  production to applications},'' \emph{Renewable and Sustainable Energy
  Reviews}, vol.~43, no.~C, pp. 1151--1158, 2015.

\bibitem{Ogungbemi2}
E.~Ogungbemi, O.~Ijaodola, F.~Khatib, T.~Wilberforce, Z.~El~Hassan,
  J.~Thompson, M.~Ramadan, and A.~Olabi, ``{Fuel cell membranes – Pros and
  cons},'' \emph{Energy}, vol. 172, no.~C, pp. 155--172, 2019.

\bibitem{Dubau3}
L.~Dubau, L.~Castanheira, F.~Maillard, M.~Chatenet, O.~Lottin, G.~Maranzana,
  J.~Dillet, A.~Lamibrac, J.~C. Perrin, E.~Moukheiber, A.~ElKaddouri,
  G.~De~Moor, C.~Bas, L.~Flandin, and N.~Caqu{\'e}, ``{A review of PEM fuel
  cell durability: materials degradation, local heterogeneities of aging and
  possible mitigation strategies},'' \emph{{WIREs Energy and Environment}},
  vol.~3, no.~6, pp. 540--560, 2014.

\bibitem{9146674}
A.~Amamou, M.~Kandidayeni, S.~Kelouwani, and L.~Boulon, ``An online self cold
  startup methodology for pem fuel cells in vehicular applications,''
  \emph{IEEE Transactions on Vehicular Technology}, vol.~69, no.~12, pp.
  14\,160--14\,172, 2020.

\bibitem{LIU2020115110}
Z.~Liu, J.~Chen, H.~Liu, C.~Yan, Y.~Hou, Q.~He, J.~Zhang, and D.~Hissel,
  ``Anode purge management for hydrogen utilization and stack durability
  improvement of pem fuel cell systems,'' \emph{Applied Energy}, vol. 275, p.
  115110, 2020.

\bibitem{Esmaili2020ModelBW}
Q.~Esmaili, M.~E. Nimvari, N.~F. Jouybari, and Y.~S. Chen, ``Model based water
  management diagnosis in polymer electrolyte membrane fuel cell,''
  \emph{International Journal of Hydrogen Energy}, vol.~45, pp.
  15\,618--15\,629, 2020.

\bibitem{4939358}
L.~{Palma} and P.~N. {Enjeti}, ``A modular fuel cell, modular dc–dc converter
  concept for high performance and enhanced reliability,'' \emph{IEEE
  Transactions on Power Electronics}, vol.~24, no.~6, pp. 1437--1443, 2009.

\bibitem{6043104}
E.~{Frappé}, A.~{De Bernardinis}, G.~{Coquery}, O.~{Bethoux}, and
  C.~{Marchand}, ``A soft-switching four-port dc-dc converter for segmented pem
  fuel cell power management in vehicle application,'' in \emph{2011 IEEE
  Vehicle Power and Propulsion Conference}, 6-9 Sept. 2011, pp. 1--6, Chicago,
  IL, USA.

\bibitem{WENG20103664}
F.~B. Weng, C.~Y. Hsu, and C.~W. Li, ``Experimental investigation of pem fuel
  cell aging under current cycling using segmented fuel cell,''
  \emph{International Journal of Hydrogen Energy}, vol.~35, no.~8, pp. 3664 --
  3675, 2010.

\bibitem{TOLJ201113105}
I.~Tolj, D.~Bezmalinovic, and F.~Barbir, ``Maintaining desired level of
  relative humidity throughout a fuel cell with spatially variable heat removal
  rates,'' \emph{International Journal of Hydrogen Energy}, vol.~36, no.~20,
  pp. 13\,105 -- 13\,113, 2011.

\bibitem{https://doi.org/10.1002/asjc.1827}
J.~Chen, Z.~Wu, C.~Wu, and C.~Yan, ``Observer based fuel delivery control for
  pem fuel cells with a segmented anode model,'' \emph{Asian Journal of
  Control}, vol.~21, no.~4, pp. 1781--1795, 2019.

\bibitem{CHEN20081179}
Y.~{S. Chen} and H.~Peng, ``A segmented model for studying water transport in a
  pemfc,'' \emph{Journal of Power Sources}, vol. 185, no.~2, pp. 1179 -- 1192,
  2008.

\bibitem{CHEN20111992}
Y.~S. Chen and H.~Peng, ``Predicting current density distribution of proton
  exchange membrane fuel cells with different flow field designs,''
  \emph{Journal of Power Sources}, vol. 196, no.~4, pp. 1992 -- 2004, 2011.

\bibitem{HONG20171565}
L.~Hong, J.~Chen, Z.~Liu, L.~Huang, and Z.~Wu, ``A nonlinear control strategy
  for fuel delivery in pem fuel cells considering nitrogen permeation,''
  \emph{International Journal of Hydrogen Energy}, vol.~42, no.~2, pp. 1565 --
  1576, 2017.

\bibitem{10.5555/3285631}
A.~van~der Schaft, \emph{L2-Gain and Passivity Techniques in Nonlinear
  Control}, 3rd~ed.\hskip 1em plus 0.5em minus 0.4em\relax Springer Publishing
  Company, London, 2018.

\bibitem{ORTEGA2002585}
R.~Ortega, A.~{van der Schaft}, B.~Maschke, and G.~Escobar, ``Interconnection
  and damping assignment passivity-based control of port-controlled hamiltonian
  systems,'' \emph{Automatica}, vol.~38, no.~4, pp. 585 -- 596, 2002.

\bibitem{WU2019595}
C.~Wu, A.~{van der Schaft}, and J.~Chen, ``Robust trajectory tracking for
  incrementally passive nonlinear systems,'' \emph{Automatica}, vol. 107, pp.
  595--599, 2019.

\bibitem{9127091}
C.~{Wu}, A.~{van der Schaft}, and J.~{Chen}, ``Stabilization of
  port-hamiltonian systems based on shifted passivity via feedback,''
  \emph{IEEE Transactions on Automatic Control}, vol.~66, no.~5, pp.
  2219--2226, 2021.

\bibitem{CASTANOS20091611}
F.~Castaños, R.~Ortega, A.~{van der Schaft}, and A.~Astolfi, ``Asymptotic
  stabilization via control by interconnection of port-hamiltonian systems,''
  \emph{Automatica}, vol.~45, no.~7, pp. 1611 -- 1618, 2009.

\bibitem{WU2020109087}
D.~Wu, R.~Ortega, and G.~Duan, ``On universal stabilization property of
  interconnection and damping assignment control,'' \emph{Automatica}, vol.
  119, p. 109087, 2020.

\bibitem{PANTELEY1998131}
E.~Panteley and A.~Loria, ``On global uniform asymptotic stability of nonlinear
  time-varying systems in cascade,'' \emph{Systems \& Control Letters},
  vol.~33, no.~2, pp. 131 -- 138, 1998.

\bibitem{MONSHIZADEH2019108527}
P.~Monshizadeh, J.~E. Machado, R.~Ortega, and A.~{van der Schaft},
  ``Power-controlled hamiltonian systems: Application to electrical systems
  with constant power loads,'' \emph{Automatica}, vol. 109, p. 108527, 2019.

\bibitem{8386663}
A.~{Yaghmaei} and M.~J. {Yazdanpanah}, ``Structure preserving observer design
  for port-hamiltonian systems,'' \emph{IEEE Transactions on Automatic
  Control}, vol.~64, no.~3, pp. 1214--1220, 2019.

\bibitem{1024334}
R.~{Ortega}, M.~W. {Spong}, F.~{G. Estern}, and G.~{Blankenstein},
  ``Stabilization of a class of underactuated mechanical systems via
  interconnection and damping assignment,'' \emph{IEEE Transactions on
  Automatic Control}, vol.~47, no.~8, pp. 1218--1233, 2002.

\bibitem{DONAIRE2016118}
A.~Donaire, R.~Ortega, and J.~Romero, ``Simultaneous interconnection and
  damping assignment passivity-based control of mechanical systems using
  dissipative forces,'' \emph{Systems \& Control Letters}, vol.~94, pp. 118 --
  126, 2016.

\bibitem{9416793}
A.~van~der Schaft and D.~Jeltsema, ``Limits to energy conversion,'' \emph{IEEE
  Transactions on Automatic Control}, vol.~67, no.~1, pp. 532--538, 2022.

\bibitem{960344}
V.~Petrovic, R.~{Ortega}, and A.~M. {Stankovic}, ``Interconnection and damping
  assignment approach to control of pm synchronous motors,'' \emph{IEEE
  Transactions on Control Systems Technology}, vol.~9, no.~6, pp. 811--820,
  2001.

\bibitem{832798}
V.~{Petrovic}, R.~{Ortega}, and A.~M. {Stankovic}, ``A globally convergent
  energy-based controller for pm synchronous motors,'' in \emph{Proceedings of
  the 38th IEEE Conference on Decision and Control}, vol.~1, 7-10 Dec. 1999,
  pp. 334--340, Phoenix, AZ, USA.

\bibitem{BENMOUNA201922467}
A.~Benmouna, M.~Becherif, J.~Chen, H.~Chen, and D.~Depernet, ``Interconnection
  and damping assignment passivity based control for fuel cell and battery
  vehicle: Simulation and experimentation,'' \emph{International Journal of
  Hydrogen Energy}, vol.~44, no.~39, pp. 22\,467 -- 22\,477, 2019.

\bibitem{HILAIRET20131097}
M.~Hilairet, M.~Ghanes, O.~Béthoux, V.~Tanasa, J.~P. Barbot, and D.~N. Cyrot,
  ``A passivity-based controller for coordination of converters in a fuel cell
  system,'' \emph{Control Engineering Practice}, vol.~21, no.~8, pp. 1097 --
  1109, 2013.

\bibitem{9032209}
P.~{Mungporn}, P.~{Thounthong}, B.~{Yodwong}, C.~{Ekkaravarodome},
  A.~{Bilsalam}, S.~{Pierfederici}, D.~{Guilbert}, B.~{N. Mobarakeh},
  N.~{Bizon}, Z.~{Shah}, S.~{Khomfoi}, P.~{Kumam}, and P.~{Burikham},
  ``Modeling and control of multiphase interleaved fuel-cell boost converter
  based on hamiltonian control theory for transportation applications,''
  \emph{IEEE Transactions on Transportation Electrification}, vol.~6, no.~2,
  pp. 519--529, 2020.

\bibitem{8818313}
S.~{Pang}, B.~{N. Mobarakeh}, S.~{Pierfederici}, M.~{Phattanasak},
  Y.~{Huangfu}, G.~{Luo}, and F.~{Gao}, ``Interconnection and damping
  assignment passivity-based control applied to on-board dc–dc power
  converter system supplying constant power load,'' \emph{IEEE Transactions on
  Industry Applications}, vol.~55, no.~6, pp. 6476--6485, 2019.

\bibitem{CHEN202048}
J.~Chen, L.~Huang, C.~Yan, and Z.~Liu, ``A dynamic scalable segmented model of
  pem fuel cell systems with two-phase water flow,'' \emph{Mathematics and
  Computers in Simulation}, vol. 167, pp. 48--64, 2020.

\bibitem{6919261}
M.~Hilairet, O.~Béthoux, M.~Ghanes, V.~Tanasa, J.-P. Barbot, and M.-D.
  Normand-Cyrot, ``Experimental validation of a sampled-data passivity-based
  controller for coordination of converters in a fuel cell system,'' \emph{IEEE
  Transactions on Industrial Electronics}, vol.~62, no.~8, pp. 5187--5194,
  2015.

\bibitem{RAKHTALA2014203}
S.~M. Rakhtala, A.~R. Noei, R.~Ghaderi, and E.~Usai, ``Design of finite-time
  high-order sliding mode state observer: A practical insight to pem fuel cell
  system,'' \emph{Journal of Process Control}, vol.~24, no.~1, pp. 203 -- 224,
  2014.

\bibitem{Utkin}
V.~Utkin, ``Discussion aspects of high-order sliding mode control,'' \emph{IEEE
  Transactions on Automatic Control}, vol.~61, no.~3, pp. 829--833, 2016.

\bibitem{doi:10.1080/00207170802590531}
J.~Davila, L.~Fridman, A.~Pisano, and E.~Usai, ``Finite-time state observation
  for non-linear uncertain systems via higher-order sliding modes,''
  \emph{International Journal of Control}, vol.~82, no.~8, pp. 1564--1574,
  2009.

\bibitem{Levant}
A.~Levant, ``Higher-order sliding modes, differentiation and output-feedback
  control,'' \emph{International Journal of Control}, vol.~76, no. 9-10, pp.
  924--941, 2003.

\bibitem{Pukrushpan_Thesis}
J.~T. Pukrushpan, ``Modeling and control of fuel cell systems and fuel
  processors,'' \emph{Ph.D. dissertation, University of Michigan}, 2003.

\end{thebibliography}
\bibliographystyle{elsarticle}


\end{document}